\documentclass{raa}            
\usepackage{graphicx}             
\usepackage{natbib}
\usepackage{amssymb,amsmath}
\begin{document}

   \title{A Method of Alignment of the Plastic Scintillator Detector of DAMPE
}

   \setcounter{page}{1}          

   \author{Peng-Xiong Ma,
      \inst{1,2}
      \and Yong-Jie Zhang
      \inst{3,4,5}
      \and Ya-Peng Zhang
      \inst{3}
      \and{Yao Li}
     \inst{3}
   \and Jing-Jing Zang
    \inst{1}
   \and Xiang Li
    \inst{1}
   \and Tie-Kuang Dong
   \inst{1,2}
           \and Yi-Zhong Fan
   \inst{1,2}
   \and Shi-Jun Lei
   \inst{1,2}
       \and Jian Wu
      \inst{1,2}
           \and Yu-Hong Yu
      \inst{3}
      \and Qiang Yuan
      \inst{1,2}
    \and Chuan Yue
    \inst{1,5}
      \and Zhi-Yu Sun
      \inst{3}
   }
   \institute{Key Laboratory of Dark Matter and Space Astronomy, Purple Mountain Observatory, Chinese Academy of Sciences,2 West Beijing Road, Nanjing 210008, China; {\it zangjj@pmo.ac.cn};  {\it xiangli@pmo.ac.cn }\\
        \and
             School of Astronomy and Space Science, University of Science and Technonlogy of China, Hefei 230026, China\\
        \and
           Institute of Modern Physics, Chinese Academy of Sciences, 509 NanChang Road, LanZhou 730000, China;{\it y.p.zhang@impcas.ac.cn}\\
           \and School of Nuclear Science and Technology, Lanzhou University, 222 South Tianshui Road, Lanzhou 730000, P.R. China \\
           \and
            University of Chinese Academy of Sciences, Yuquan Road 19, Beijing 100049, China \\
   }

   \date{Received~~2017 11 day; accepted~~2017~~month day}

\abstract{
The Plastic Scintillator Detector (PSD) of the DArk Matter Particle Explorer (DAMPE) is designed to measure cosmic ray charge (Z) and to act as a veto detector for gamma-ray identification. In order to fully exploit the charge identification potential of the PSD and to enhance its capability to identify the gamma ray events, we develop a PSD detector alignment method. The path length of a given track in the volume of a PSD bar is derived taking into account the shift and rotation alignment corrections. By examining energy spectra of corner-passing events and fully contained events, position shifts and rotations of all PSD bars are obtained, and found to be on average about 1mm and 0.0015 radian respectively. To validate the alignment method, we introduce the artificial shifts and rotations of PSD bars in the detector simulation. These shift and rotation parameters can be recovered successfully by the alignment procedure. As a result of the PSD alignment procedure, the charge resolution of the PSD is improved from $4\%$ to $8\%$ depending on the nuclei. 
}

   \authorrunning{P.-X. Ma et al.}            
   \titlerunning{Alignment for PSD of DAMPE }  

   \maketitle

\section{Introduction}           
\label{sect:intro}

\begin{figure}
\centering
\includegraphics[width=9cm]{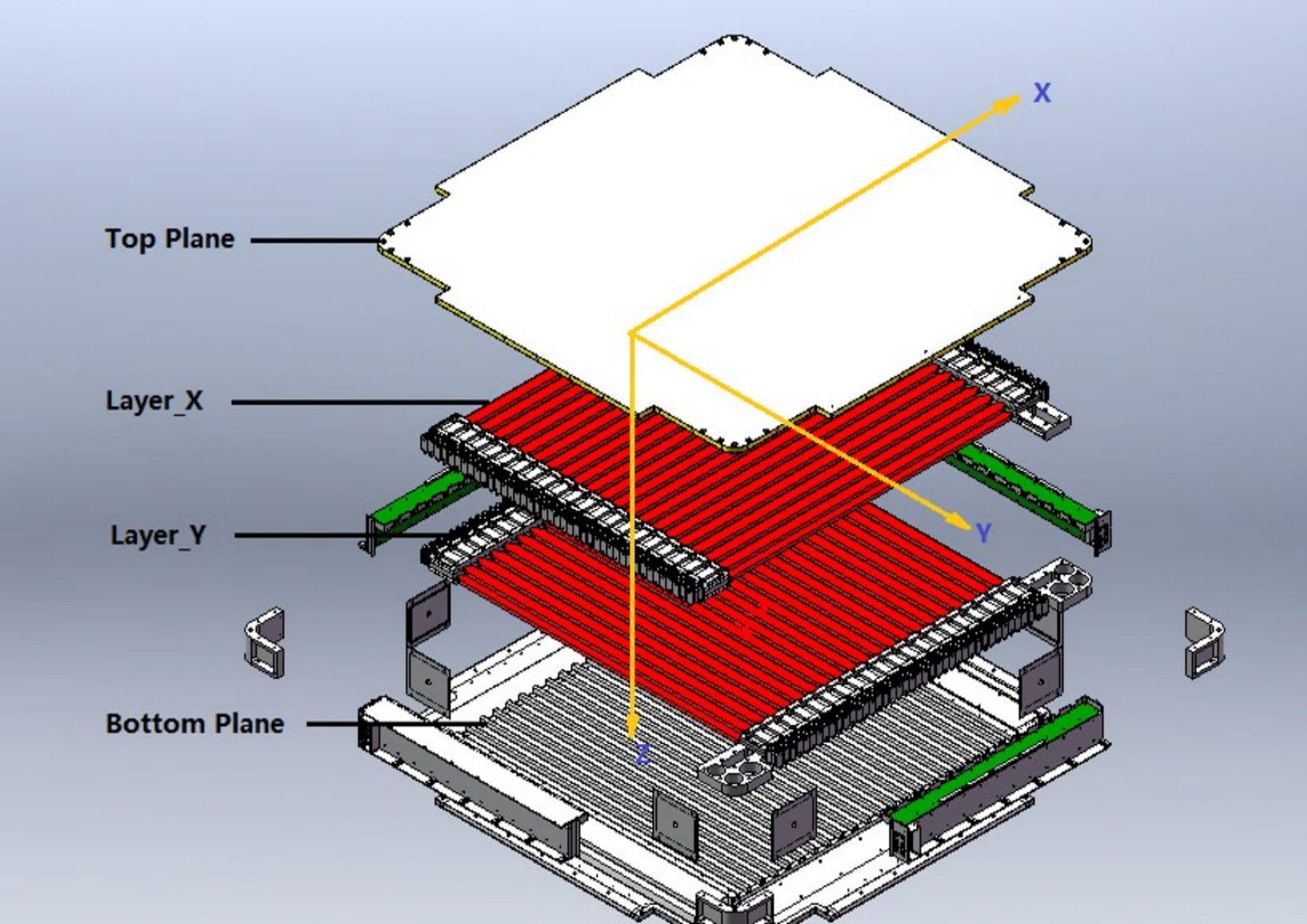}
\includegraphics[width=9cm]{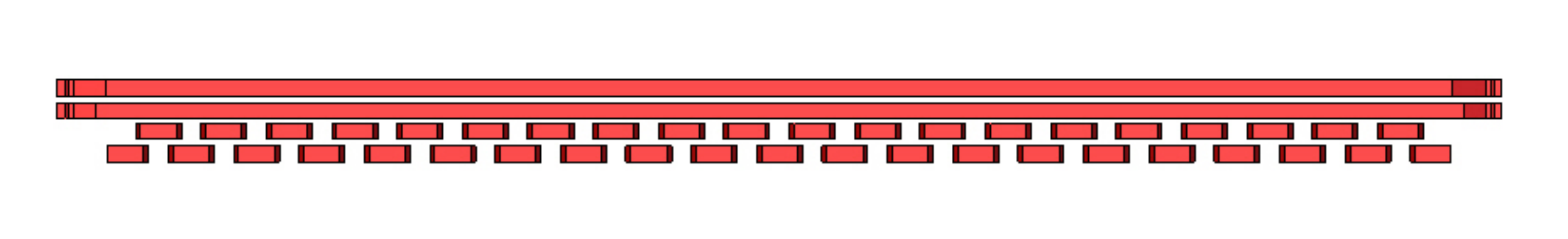}
\caption{The arrangement of PSD bars and the side view of PSD bars.}
\label{fig1}
\end{figure}

The DAMPE is a space-borne satellite of China that operates in solar synchronous orbit at an altitude of 500 km for more than two years. The payload of DAMPE is a high-energy cosmic ray detection system equipped with four sub-detectors (Chang. et al. \cite{chang17}): a Plastic Scintillator Detector (PSD) (Yu. et al. \cite{yu}), a Silicon Tungsten tracKer-convertor (STK) (P. Azzarello et al. \cite{azz16}), a BGO calorimeter (BGO) (Zhang. et al \cite{zhang16}), and a NeUtron Detectors (NUD)(He. et al. \cite{he2016}). With this design, DAMPE can measure the charge, the energy and the incoming direction of the cosmic rays. The PSD, as a thin material detector, is designed to detect the charge of the cosmic ray via measuring its energy deposition in plastic scintillator and also serves as a veto detector to identify gamma-rays from charged particles. The STK mounted below the PSD is a silicon-strip tracker with 3 layers of thin tungsten plates inserted below the first, second and third detection layer. By this design, the high-energy gamma-rays can be converted into $e^+/e^-$ pairs and then their trajectories can be reconstructed. STK is also designed to reconstruct the trajectories and measure the absolute charge (Z) of cosmic-ray ions. The BGO is a 3D imaging absolute-absorption calorimeter, which is designed to measure the energy of electrons and gamma-rays from few GeV to 10 TeV and the energy of cosmic ray nuclei from 10 GeV/n to about 200 TeV/n (DAMPE Collaboration \cite{Ele}). The bottom sub-detector of DAMPE is NUD, designed to enhance e/p separation power by detecting neutrons generated by hadronic shower in the BGO. 
The PSD is composed by 82 plastic scintillator bars arranged into two layers, both layers have 39 bars with a size of $ 824 \times 28 \times 10$ mm$^3 $ and two edge bars with a size of $ 824 \times 25 \times 10$ mm$^3$. Two layers are orthogonal to each other. In order to avoid dead regions, neighbor bars in each layer are staggered by 10 mm as shown in Fig. \ref{fig1}. The other details of the PSD  detector structure can be found in (Yu. et al. \cite{yu}). 

The mean energy deposition (or most probable value of the energy deposition) of a high-energy charged particle in a PSD bar is proportional to its path length (hereafter PL) in the volume of a PSD bar. Therefore, in order to obtain an accurate energy deposition of a charged particle in the PSD, it is important to carry out a detector alignment of all PSD bars. 
If a PSD bar is not located in its designed position, the measured energy spectrum of minimum ionization particles (MIPs) features a distorted structure due to an incorrect calculation of the PL, especially for the particles that pass only through a corner (corner-passing events). 
Based on this fact we develop a method to align all PSD bars using the correlation between measured energy spectra and PL. 
In the paper, we will firstly introduce a method of PSD alignment in Section \ref{sec2}. The validation of this method and the charge resolution improvement are presented in Section \ref{sec3}. The results and the possible application of the alignment method are presented in Section \ref{sec4}.
\section{Methodology}
\label{sec2}

As mentioned above, the energy deposition of a charged particle in a PSD bar is sensitive to its PL. Position shift or rotation of a PSD bar would cause an incorrect calculation of the PL of the charged particles and thus the measured energy spectra of MIPs may be distorted. Typically, 6 independent variables are needed to describe the position change of one PSD bar, which are three rotation angles ($\theta_{yz},\theta_{xz},\theta_{xy}$) and three shift distances ($\Delta_{x},\Delta_{y},\Delta_{z}$).

\begin{figure}
\centering
\includegraphics[width=9cm]{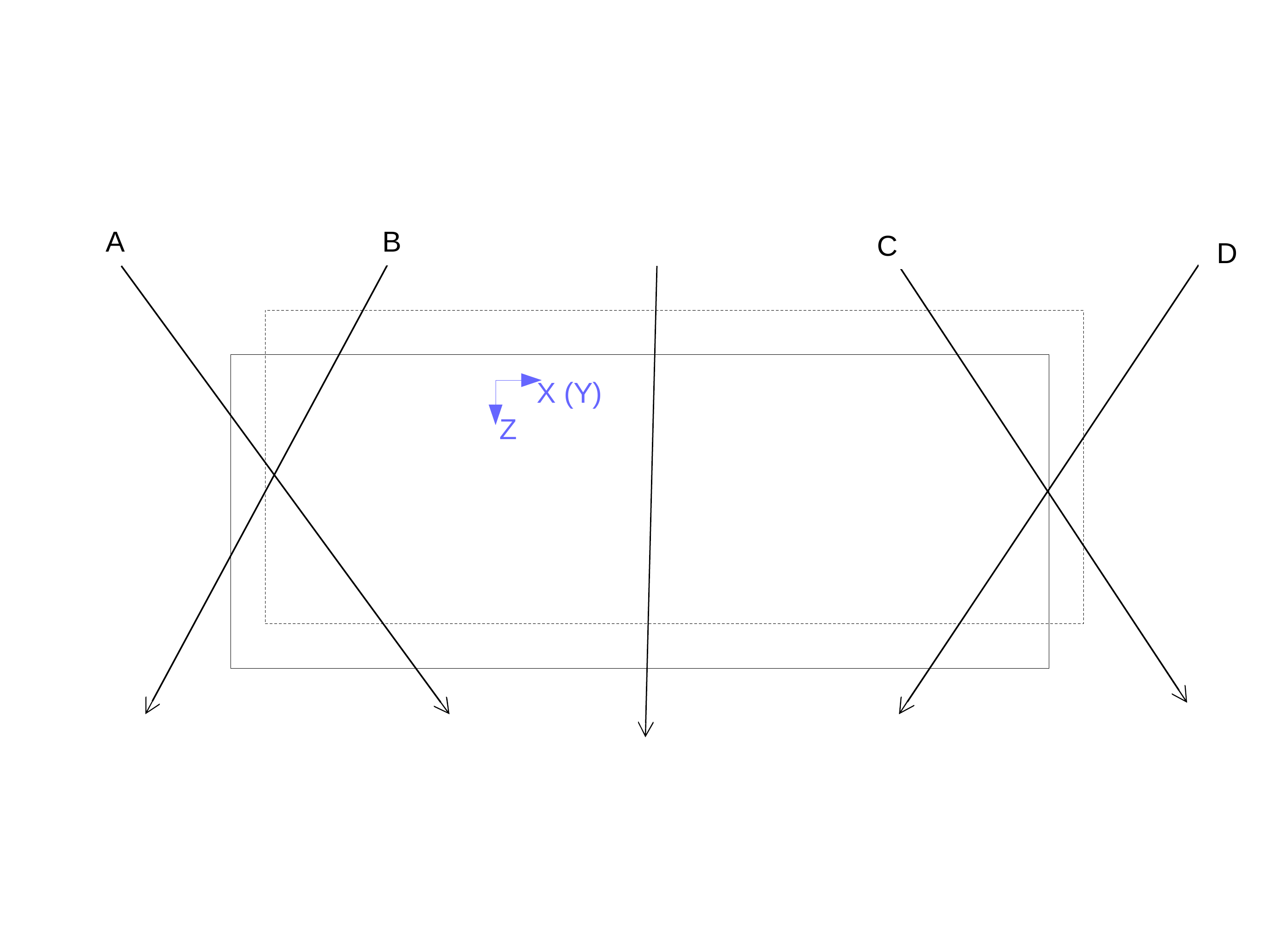}
\caption{Schematic view of misaligned PSD bar. The dashed and dotted rectangle represent the expected and real position of a PSD bar respectively. The four event types (A, B, C and D) passing through a corner of PSD bar can be used to correct for misalignment, thanks to the dependence of PL on bar misalignment.}
\label{fig2}
\end{figure}

Due to the stability of mechanical structure of DAMPE, the shifts and rotation of PSD bars are quite tiny so that can be treated as a first-order term. The events which cross the upper and lower surfaces of a PSD bar (fully contained) are not sensitive to the shift along the bar. At the same time, the fully contained events are distorted only weakly by shift and rotation (see in Fig. \ref{fig2}). Hereafter we refer to the fully contained events and their measured energy spectrum as "middle events" and "standard spectrum"

\begin{figure}
\centering
\includegraphics[width=10cm]{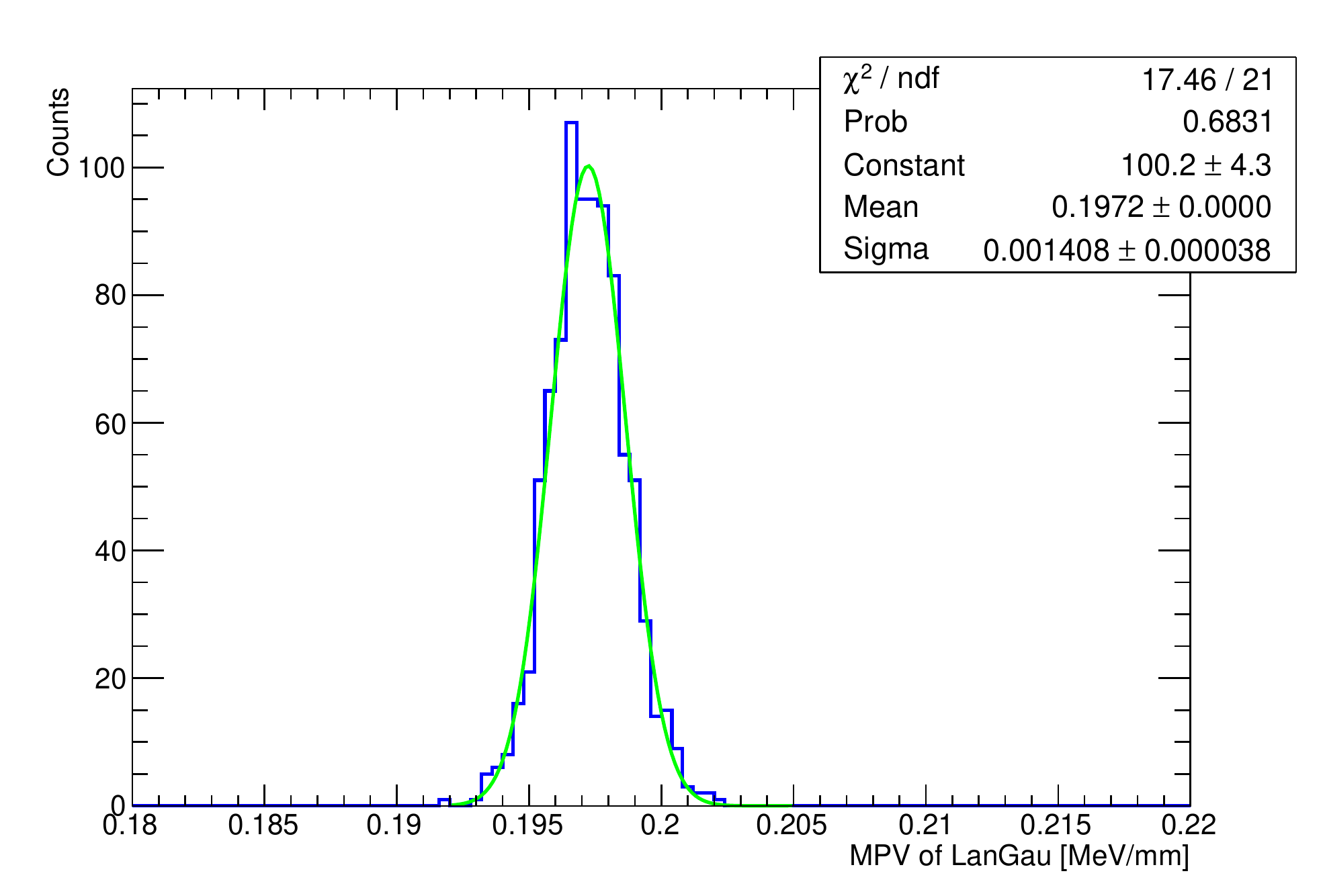}
\caption{MPV distribution of 902 segments of all PSD bars, where the MIPs events are limited to pass in the middle region of PSD bar in Fig. \ref{fig2}. It is credible that these events are affected negligibly by $H$, $V$, and $\theta_{xy}$.}
\label{fig3}
\end{figure}

As shown in Fig. \ref{fig2}, the events which are crossing a corner of a PSD bar defined as the "corner-passing events" with four cases: A, B, C and D. To convert the vertical rotation problem into shifting problem, each physical plastic bar is divided lengthwise into 11 equal virtual segments. Fig. \ref{fig3} shows the MPV distribution of the middle events in the 902 segments (11 segments for each PSD bar, 82 bars in total). As seen from the figure, deviations of the MIP spectra between the 902 segments are minor enough, meaning that the rotation angles $\theta_{yz}$ and $\theta_{xz}$ of PSD bars are negligible. In the meanwhile, shifts along the bar would not worsen the charge resolution. Finally, three effective variables remain in our alignment method, which are $\Delta_{x/y}$, $\Delta_{z}$ and $\theta_{xy}$, hereafter written as $H$, $V$ and $\theta_{xy}$.

According to the Bethe-Bloch formula, the energy deposition is proportional to the PL for a given charged particle. The track direction precision of is crucial in order to get a proper PL. According to the Geant4-based (Geant4 Collaboration \cite{geant2003} et. al.) simulation of DAMPE, the track angular resolution is about 0.0036 degree, and MIP events have a clear track in DAMPE, with almost no backscattered particles. Moreover, the interaction of MIP is a purely electromagnetic process, and therefore is well modelled in Geant4, justifying the choice of MIP events for the alignment analysis. In particular, we select MIP events according to the following criteria:

1): There is exactly one track in the event;

2): The track should have 4 $XY$ points;

3): Total number of hits in PSD is more than 0 and less than 4;

4): There are less than 3 hits in each PSD layer;

5): In the first three layers of BGO there are less than 2 hits per layer;

6): The total energy in the first three layers of BGO is less than 5 times BGO MIP energy (22.5 $MeV$).

With this selection, we can obtain about $45000$ MIP events using one-day flight data.
The deposited energy in one PSD bar, $E_{dep}$, is expressed as follows:
\begin{equation}
E_{dep} = S \cdot PL(H,V,\theta_{xy}),
\label{equ1}
\end{equation}
where $E_{dep}$ is the deposited energy in one bar, $PL(H,V,\theta_{xy})$ is the PL as a function of the alignment variables ($H,V,\theta_{xy}$), and $S$ is the most probable value of deposited energy per millimeter of fully contained events in Fig.\ref{fig2}, which is treated as the "standard value". 
At the same time, we define the deposited energy per millimeter (MeV/mm) as $\eta$ in this paper. 
As mentioned above, three alignment parameters need to be calibrated: horizontal shift ($H$), vertical shift ($V$) and rotation angle in XY plane ($\theta_{xy}$). The track of a charged particle is a 3-dimensional line given by the STK. It is defined by a space point $(P_x,P_y,P_z)$ and a direction vector $(D_x,D_y,D_z)$. The angle between track and the Z axis is defined as $\theta = \arctan\left(\sqrt{D_x^2+D_y^2}/D_z\right)$ 

If a PSD bar has a shift or rotation, the real PL is different from the ideal PL and the real PL can be calculated as
\begin{equation}
\begin{split}
PL(H,V,\theta_{xy}) =& \frac{1}{\cos\theta}\cdot\left(a\frac{D_z}{D_{x(y)}}H-aV+a\frac{D_z}{D_{x(y)}}\Delta L_i\theta_{xy}-P_z\right. \\
&\left.+a\frac{D_z}{D_{x(y)}}\left(X_0(Y_0)+b\frac{W}{2}-P_{x(y)}\right)-aZ_0+\frac{T}{2}\right),
\end{split}
\label{equ10}
\end{equation}
where $\Delta L_i$ is the offset along the bar of the $i$-th segment with respect to the center of a bar, ($X_0$, $Y_0$, $Z_0$) is the ideal geometrical center point of one PSD bar, $T$ and $W$ are respectively the thickness and the width of a PSD bar. $a=-1$ for the cases A and D see Fig. \ref{fig2} and $a=1$ for the cases B and C; $b=-1$ is for the cases A and B,  and $b=1$ for the cases C and D.

After putting Eq.~(\ref{equ10}) to Eq.~(\ref{equ1}), we get for each corner event
\begin{equation}
H - \frac{D_{x(y)}}{D_z}V + \Delta L_i\theta_{xy} = C,
\label{equ12}
\end{equation}
where
\begin{equation}
C = \frac{D_{x(y)}}{D_z}\left(\frac{aE_{dep}\cos\theta}{S} + Z_0 + aP_z-a\frac{T}{2}\right) + P_{x(y)} - X_0(Y_0) - b\frac{W}{2}.
\end{equation}

For all corner events we get the following matrix equation
\begin{equation}
\left(
\begin{matrix}
1 & \left[-\frac{D_{x(y)}}{D_z}\right]_1 & \left[\Delta L_i\right]_1 \\
1 & \left[-\frac{D_{x(y)}}{D_z}\right]_2 & \left[\Delta L_i\right]_2 \\
\vdots & \vdots & \vdots \\
1 & \left[-\frac{D_{x(y)}}{D_z}\right]_N & \left[\Delta L_i\right]_N \\
\end{matrix}
\right)
\left(
\begin{matrix}
H \\
V \\
\theta_{xy}\\
\end{matrix}
\right)
=
\left(
\begin{matrix}
\left[C\right]_1 \\
\left[C\right]_2 \\
\vdots\\
\left[C\right]_N \\
\end{matrix}
\right).
\label{equ11}
\end{equation}
Where $N$ is the number of corner events in the alignment data sample and the matrix has a least square solution for $(H, V, \theta_{xy})$.

We iteratively look for the least square solution of the above matrix equation for $(H,V,\theta_{xy})$, as follows:

{\bf step 1}: construct the matrix from the corner-passing events, and then, find a least squares solution;

{\bf step 2}: use $(H, V, \theta_{xy})$ to calculate the aligned geometry;

{\bf step 3}: 
use the aligned geometry to construct the matrix (Eq.~\ref{equ11}) and get the updated least square solution $(H + \delta H, V + \delta V, \theta_{xy} + \delta\theta_{xy})$;

{\bf step 4}: 
repeat steps 1-3 until  $|\delta H| < 1~\mathrm{\mu m}, |\delta V| < 1~\mathrm{\mu m}$ and $|\delta\theta_{xy}| < 10~\mathrm{\mu rad}$.

Every variable is set at zero before alignment, after less than about 15 iterations we get the final alignment constants $(H; V; \theta_{xy})$, the most significant convergences always appear in the first step, as seen in Fig.~\ref{fig9}

\section{Validation of Alignment and Improvement of charge resolution}
\label{sec3}
\begin{figure}
\includegraphics[width=7.5cm]{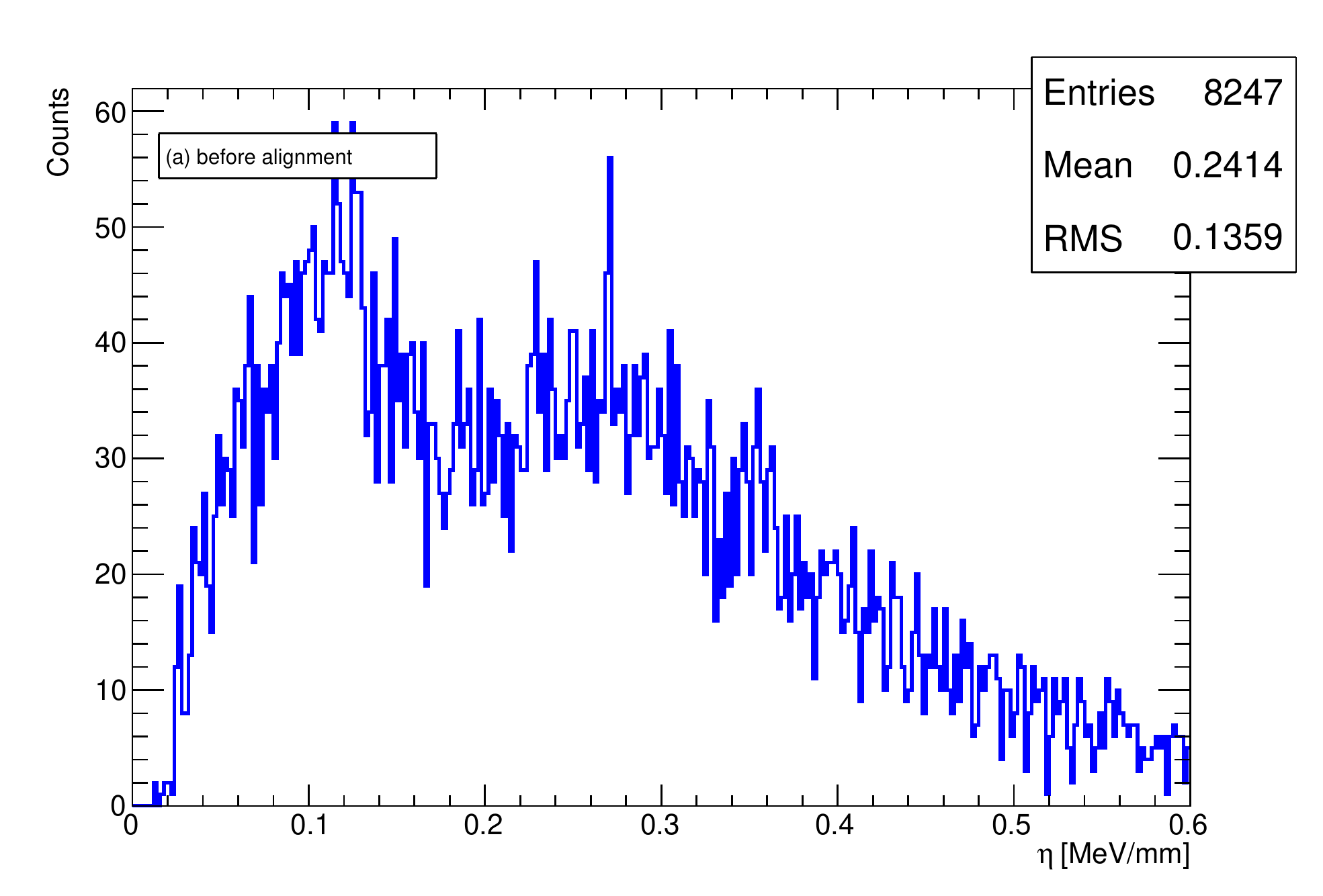}
\includegraphics[width=7.5cm]{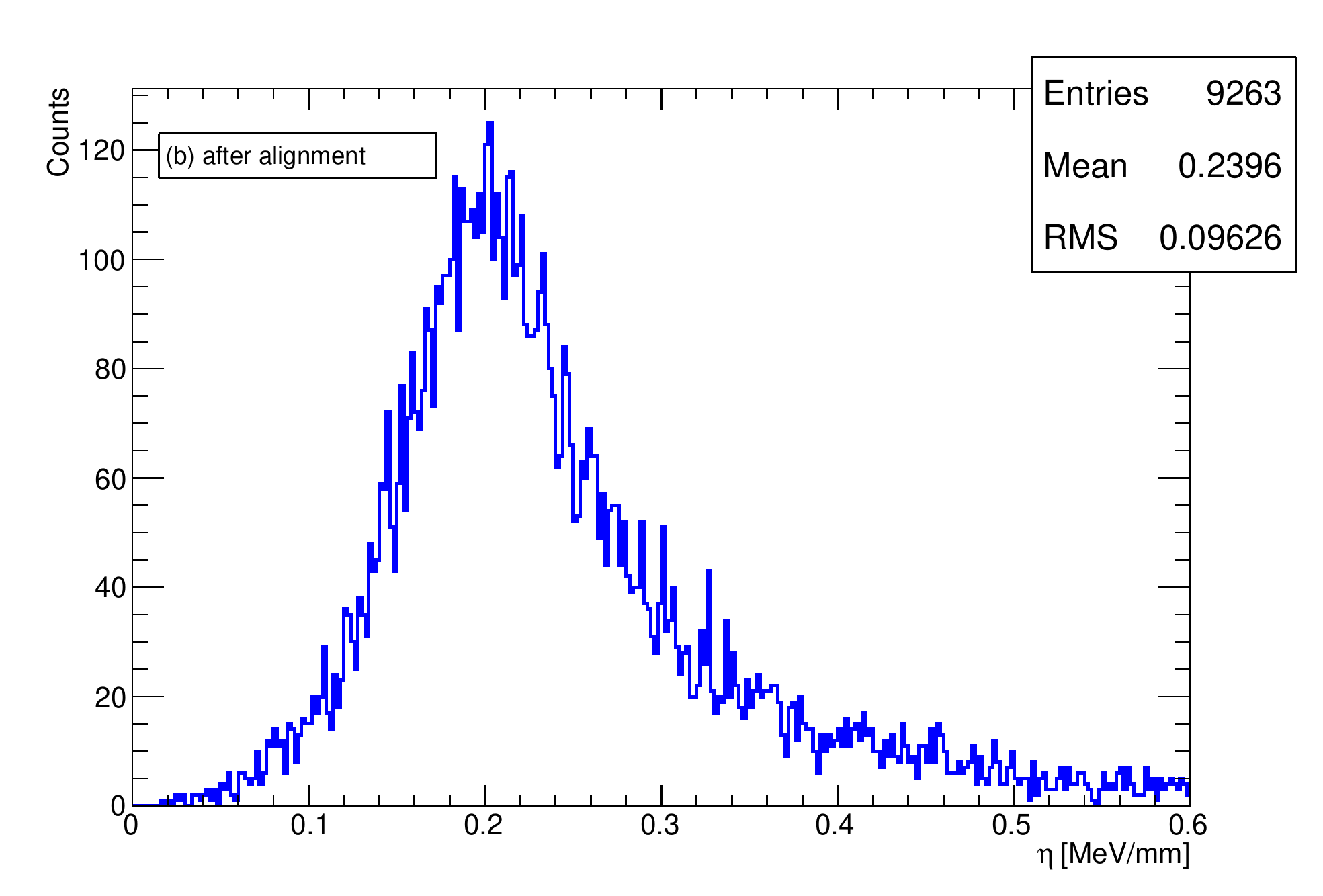}
\includegraphics[width=7.5cm]{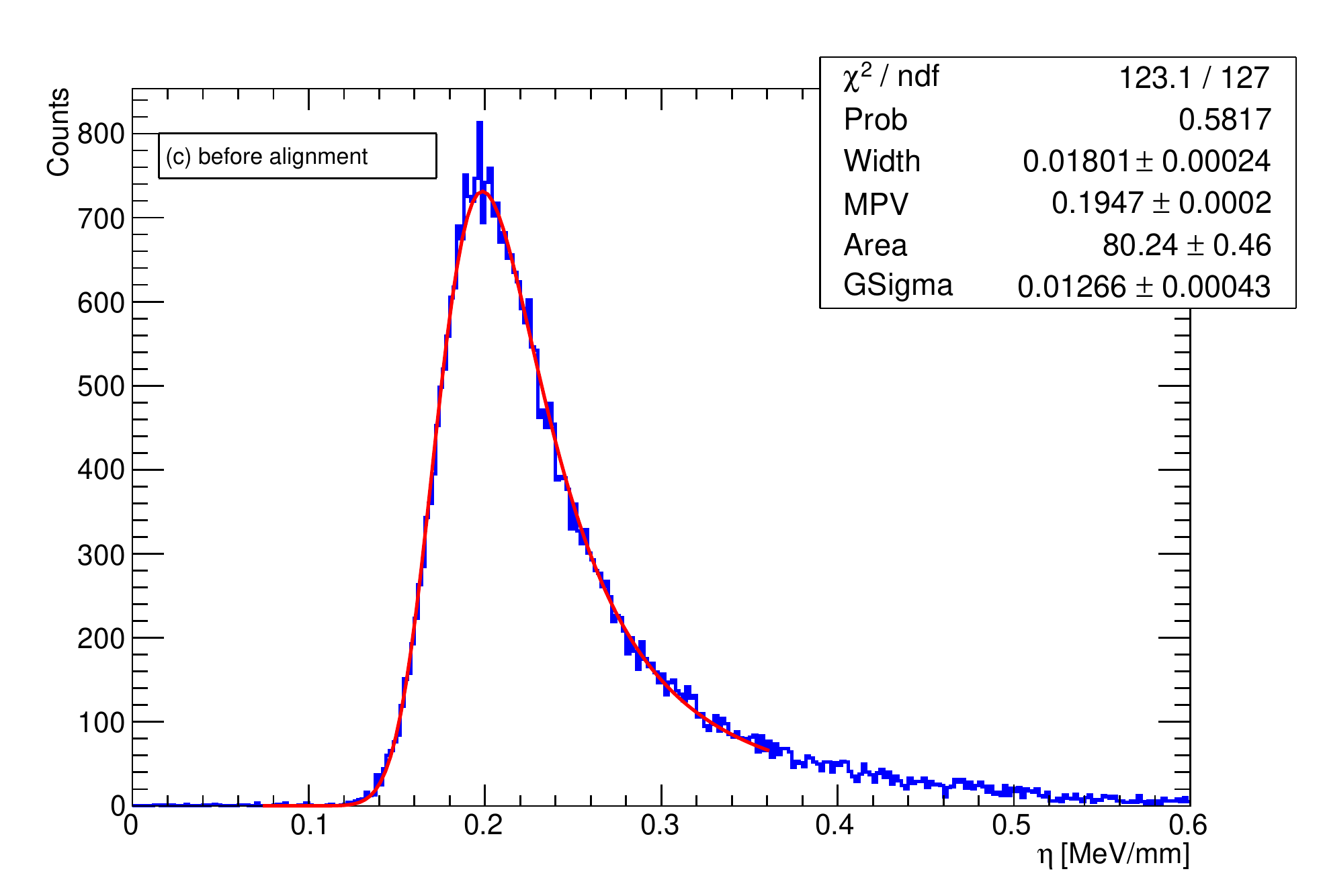}
\includegraphics[width=7.5cm]{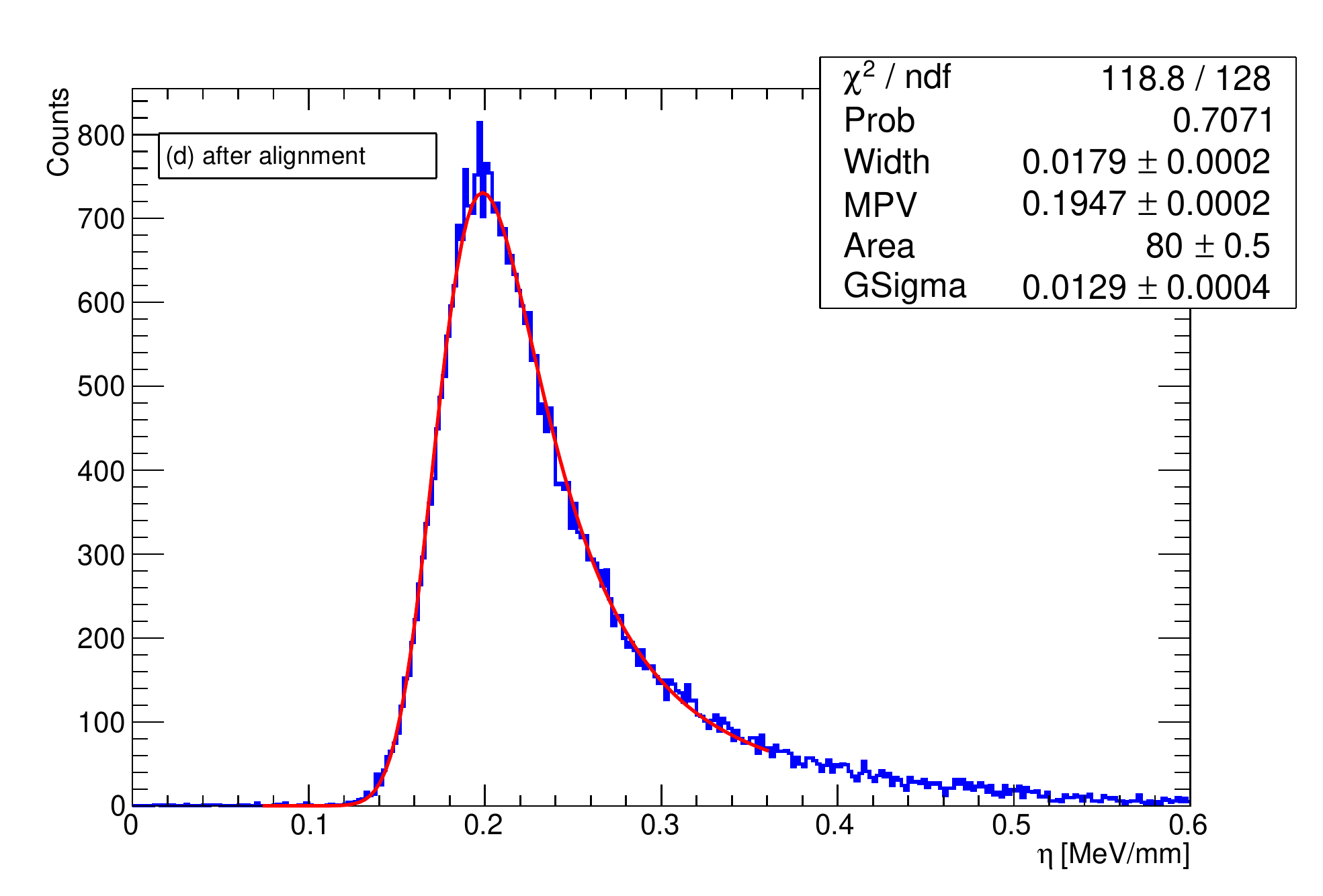}
\caption{The $\eta$ distribution of MIP events passing through the 6th segment of the 23rd PSD bar in the first layer. Four cases are shown: corner events before (a) and (b) after the alignment and middle events before (c) and after (d) the alignment. The red lines correspond to the fit with Landau convoluted with a Gaussian function.}
\label{fig4}
\end{figure}

Due to the shift and rotation, the path length of MIP events within a PSD bar will be calculated incorrectly, if no alignment is preformed. As a result, a double-peak structure in $\eta$ distribution of corner events is observed in all PSD bars. Fig. \ref{fig4}(a) shows a typical result for a PSD bar that also demonstrates that precise alignment is required. The $\eta$ distributions in Fig. \ref{fig4} are fitted with a Landau convoluted with a Gaussian function. Based on the alignment methodology in Section \ref{sec2}, $\eta$ is re-calculated iteratively. Results indicate the double peak structures of corner events are eliminated significantly and the charge resolution of proton MIP events is improved about 1.3 times, as shown in Fig. \ref{fig4}(b). Meanwhile, the change of $\eta$ spectra of middle-region events before and after the PSD alignment is minor. Fit results in Fig. \ref{fig4}(c) and (d) show that the $\eta$ spectrum of the middle events does not improve after the alignment.

\begin{figure}
\includegraphics[width=4.5cm]{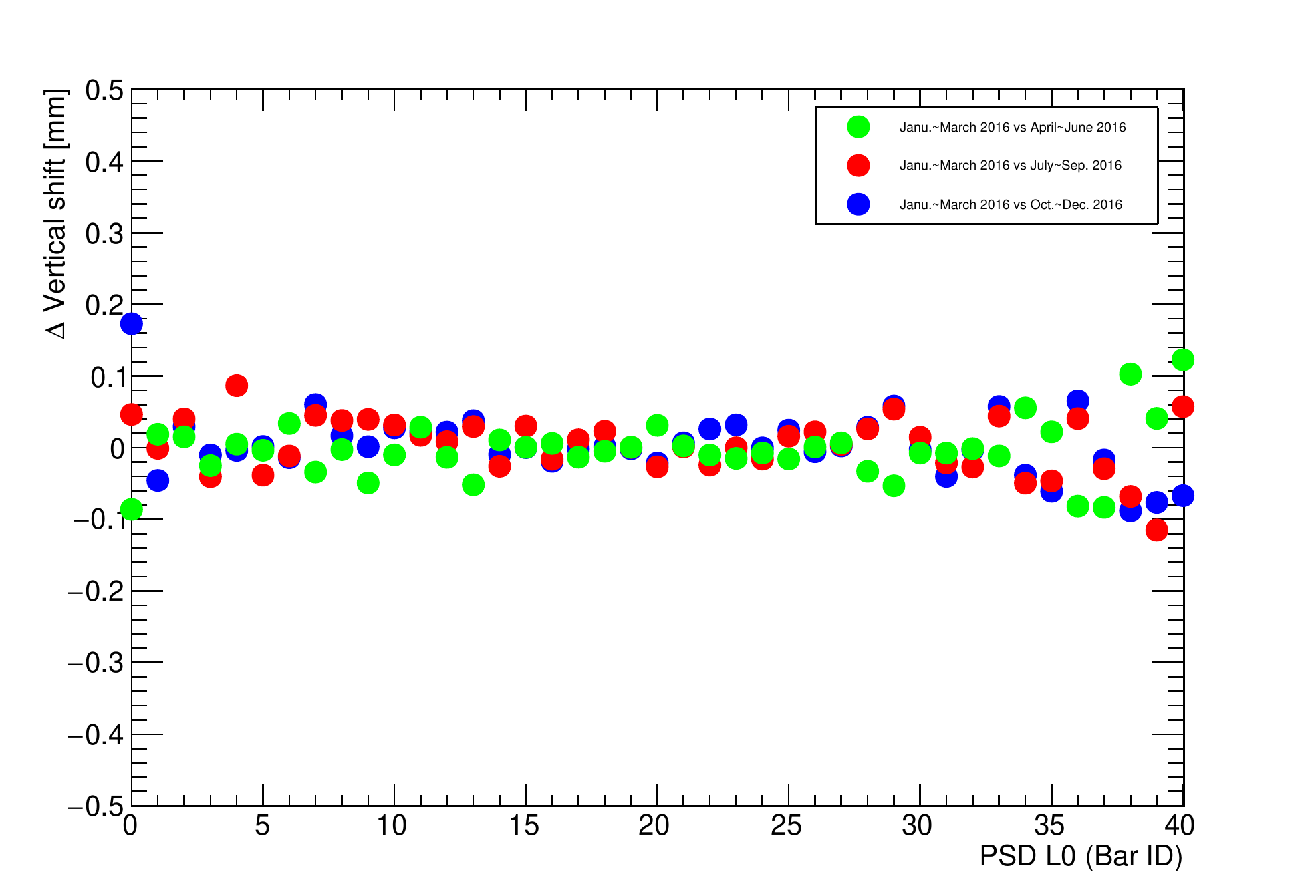}
\includegraphics[width=4.5cm]{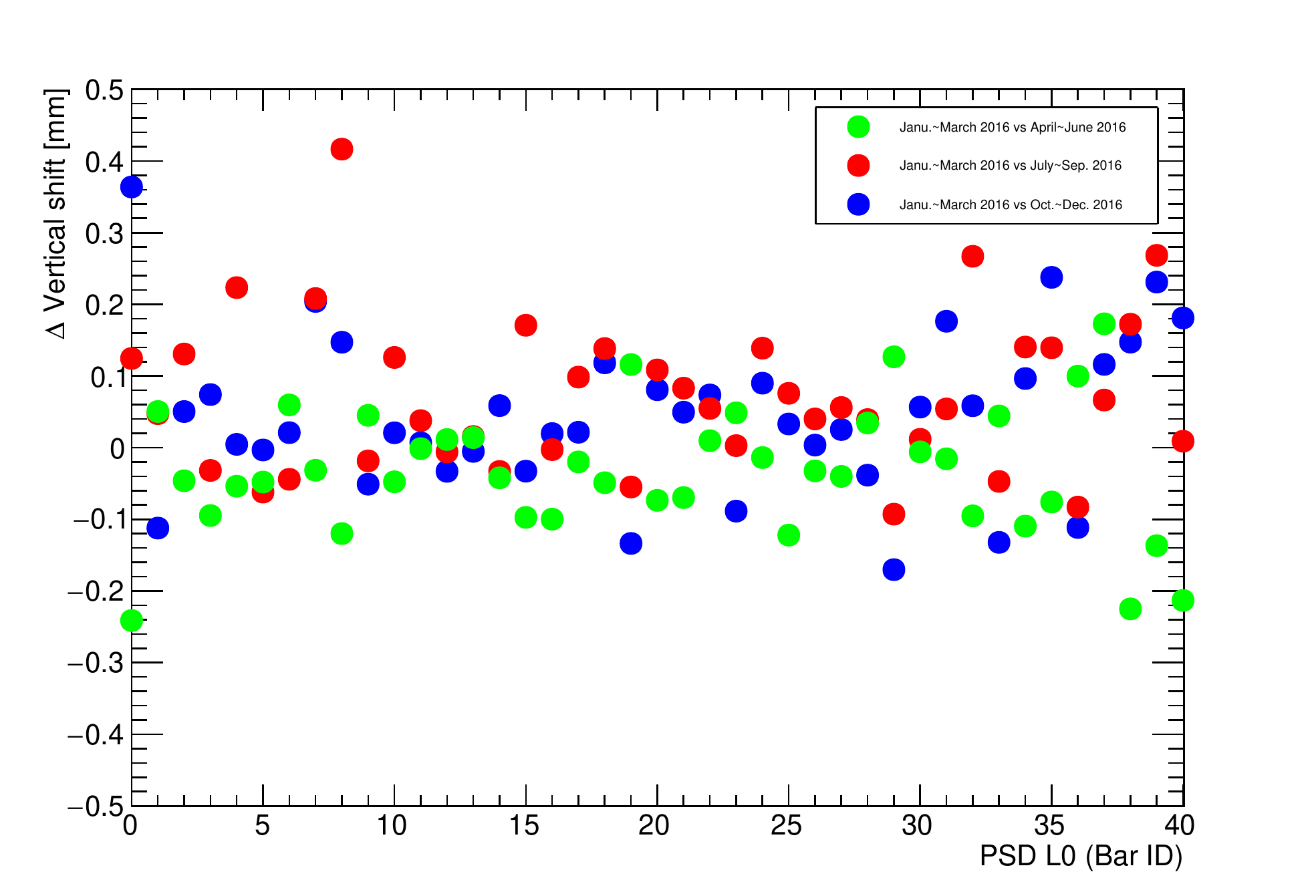}
\includegraphics[width=4.5cm]{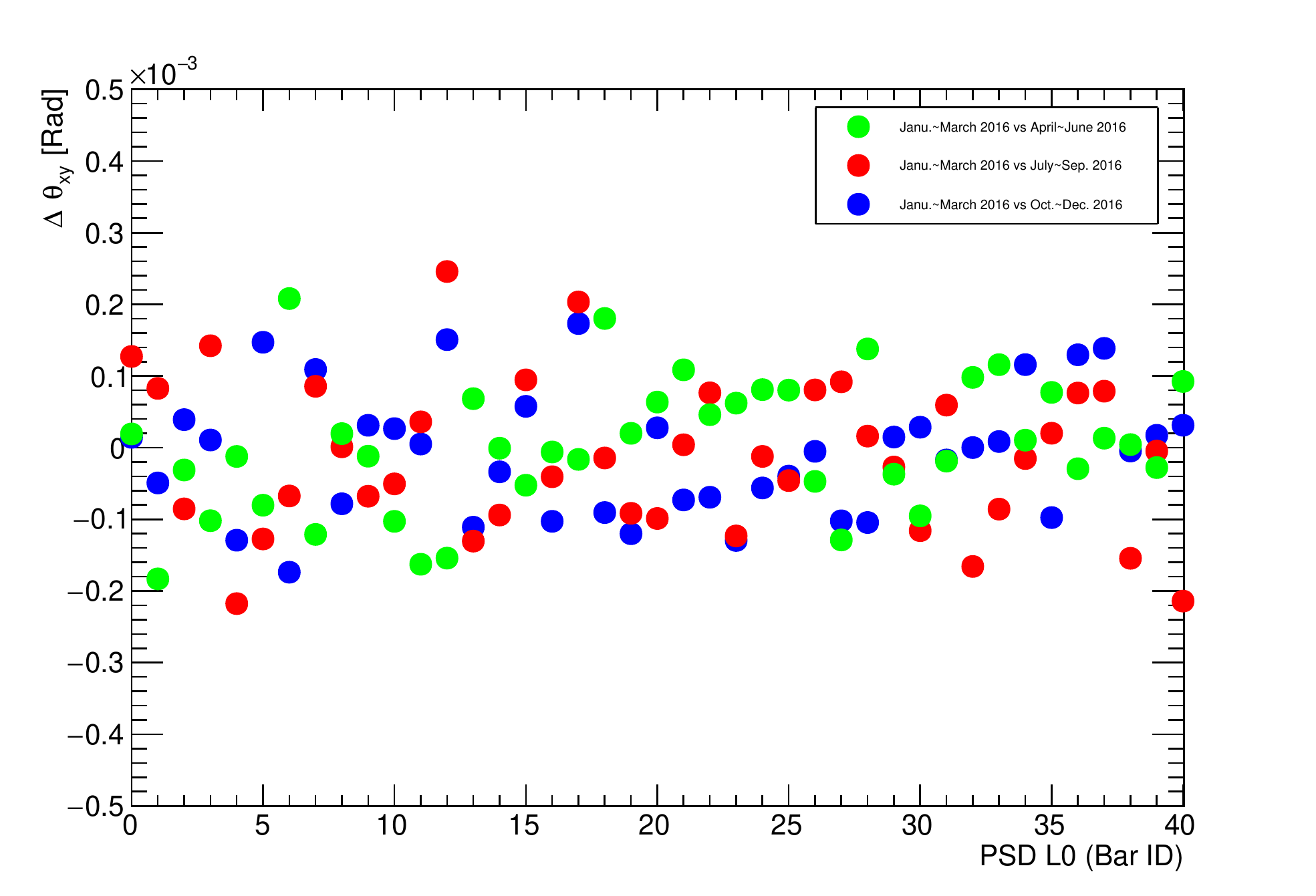}\\
\includegraphics[width=4.5cm]{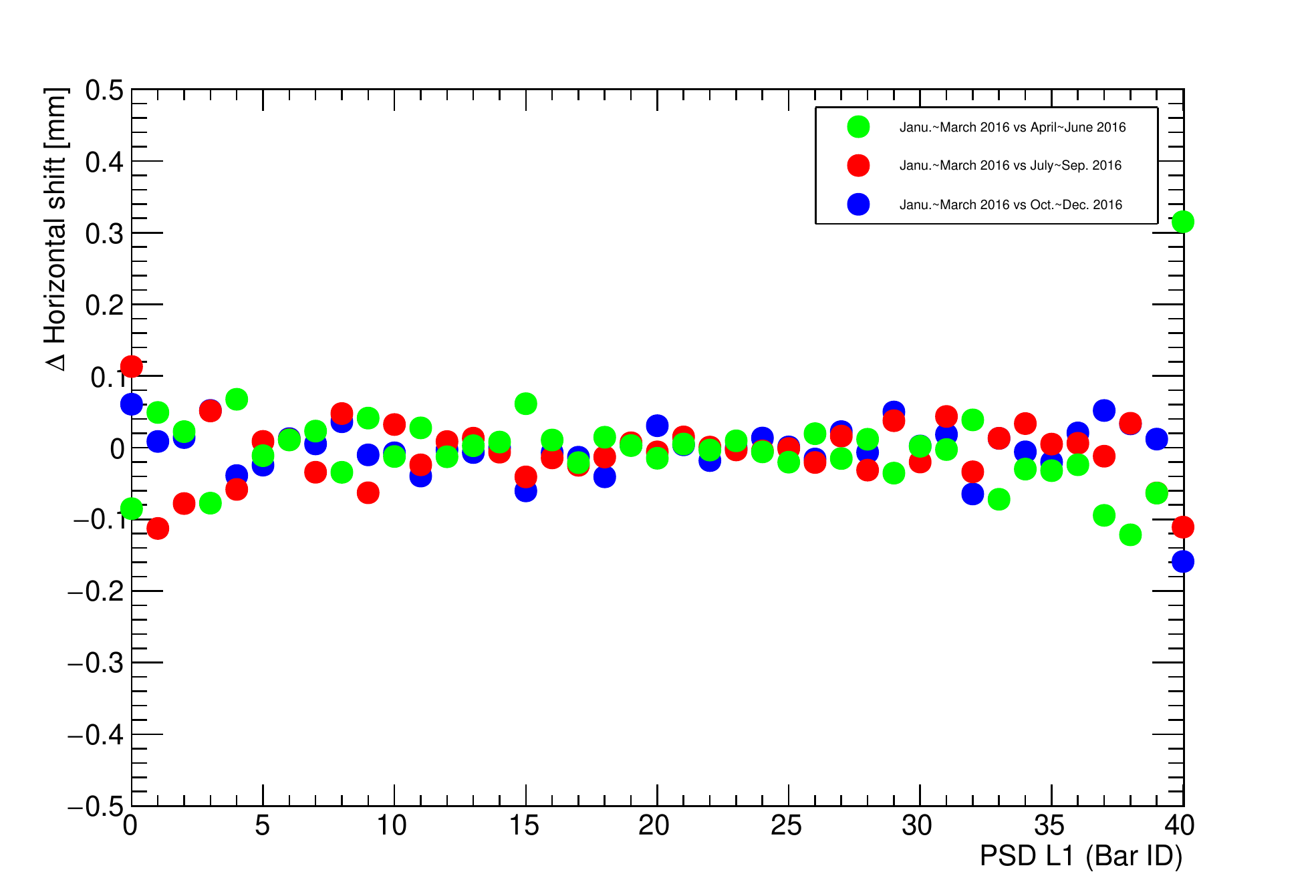}
\includegraphics[width=4.5cm]{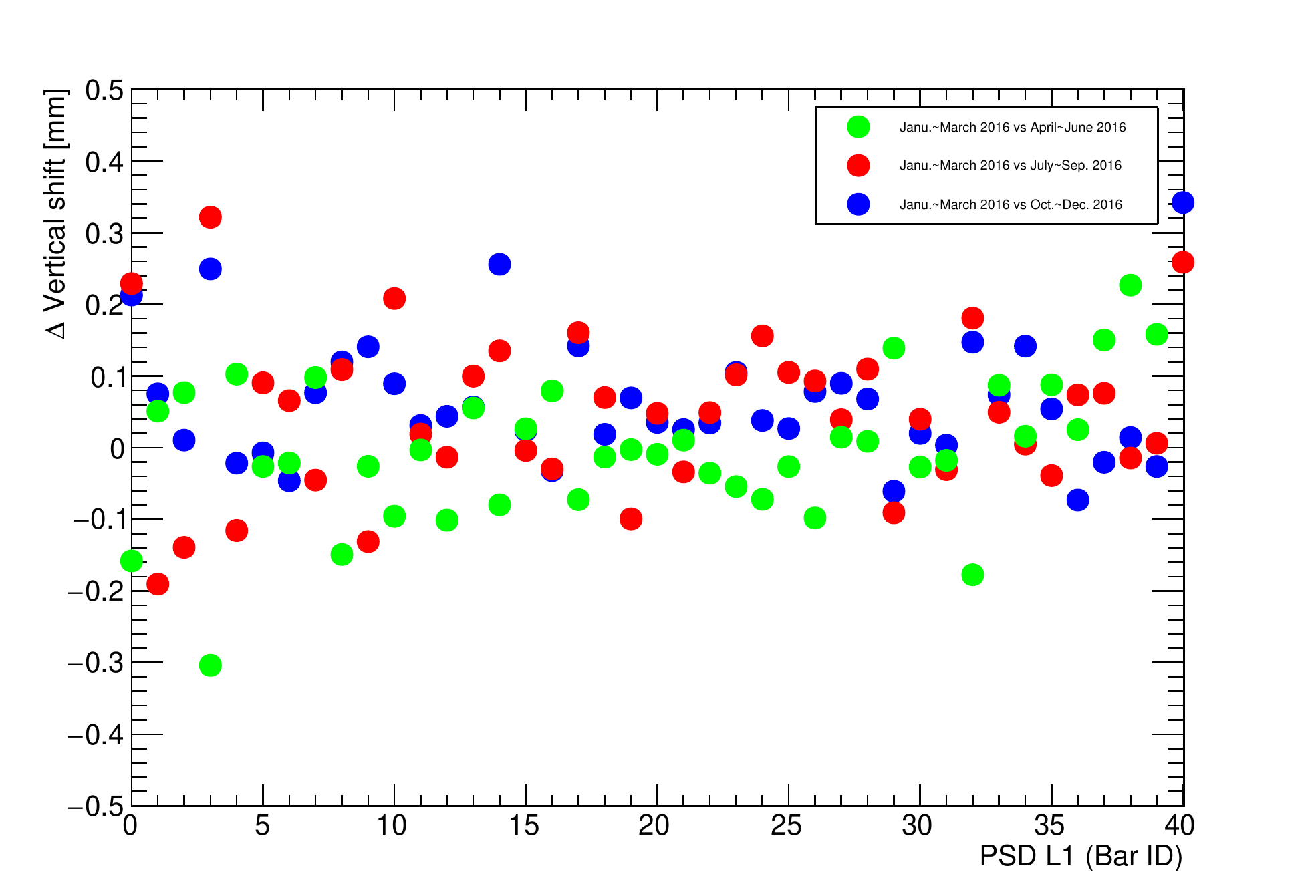}
\includegraphics[width=4.5cm]{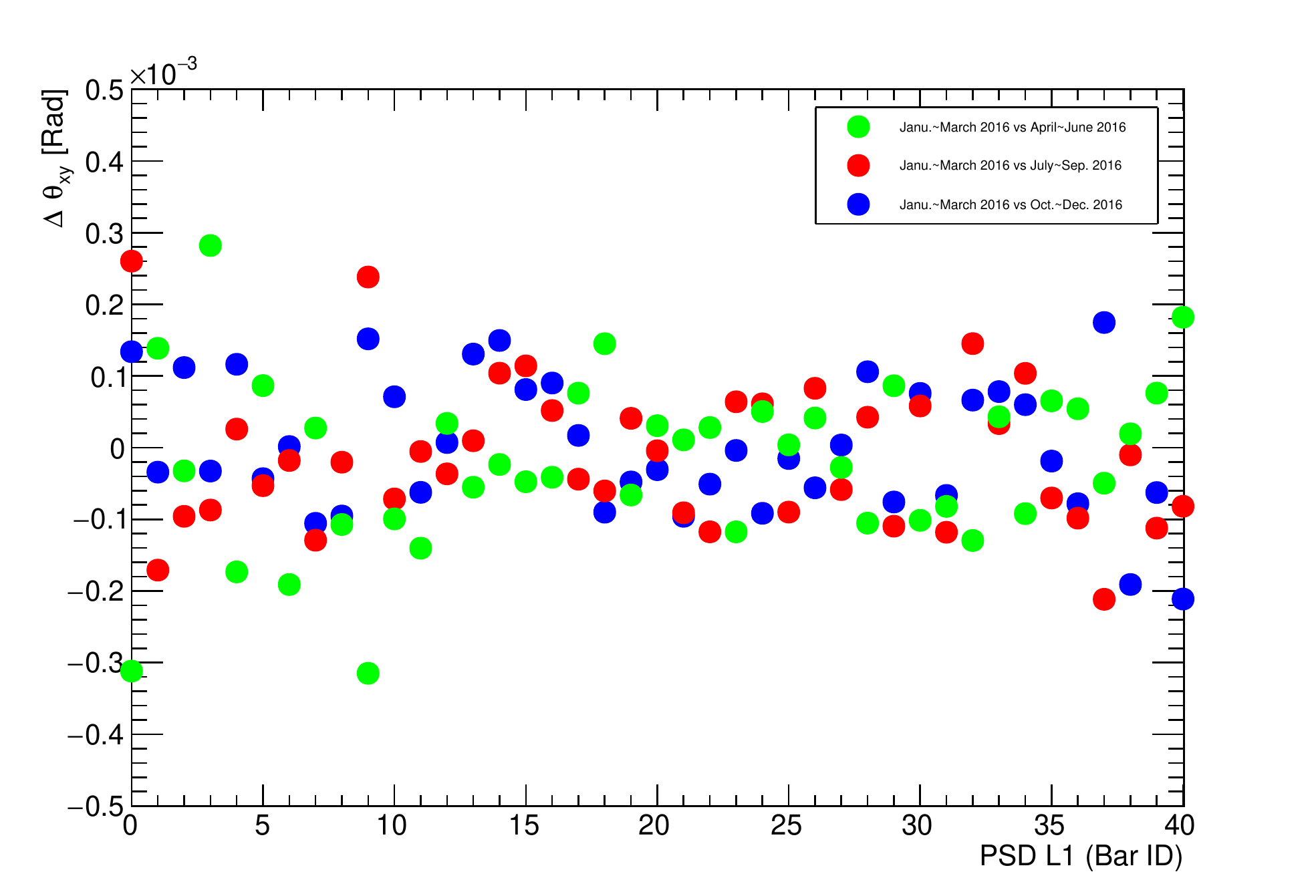}

\caption{
Variations of misalignment parameters in different time. 
Each filled circle represents one PSD bar, the green ones show variations between Jan. 2016 - Mar. 2016 and Apr. 2016 - June 2016, the red ones show variations between Jan. 2016 - Mar. 2016 and July 2016 - Sep. 2016, the blue ones show variations between Jan. 2016 - Mar. 2016 and Oct. 2016 - Dec. 2016. The two rows are for PSD first layer and second layer, respectively. The three columns from left to right correspond to variations of $H$, $V$ and $\theta_{xy}$, respectively.}
\label{fig5}
\end{figure}

\begin{figure}
\includegraphics[width=4.5cm]{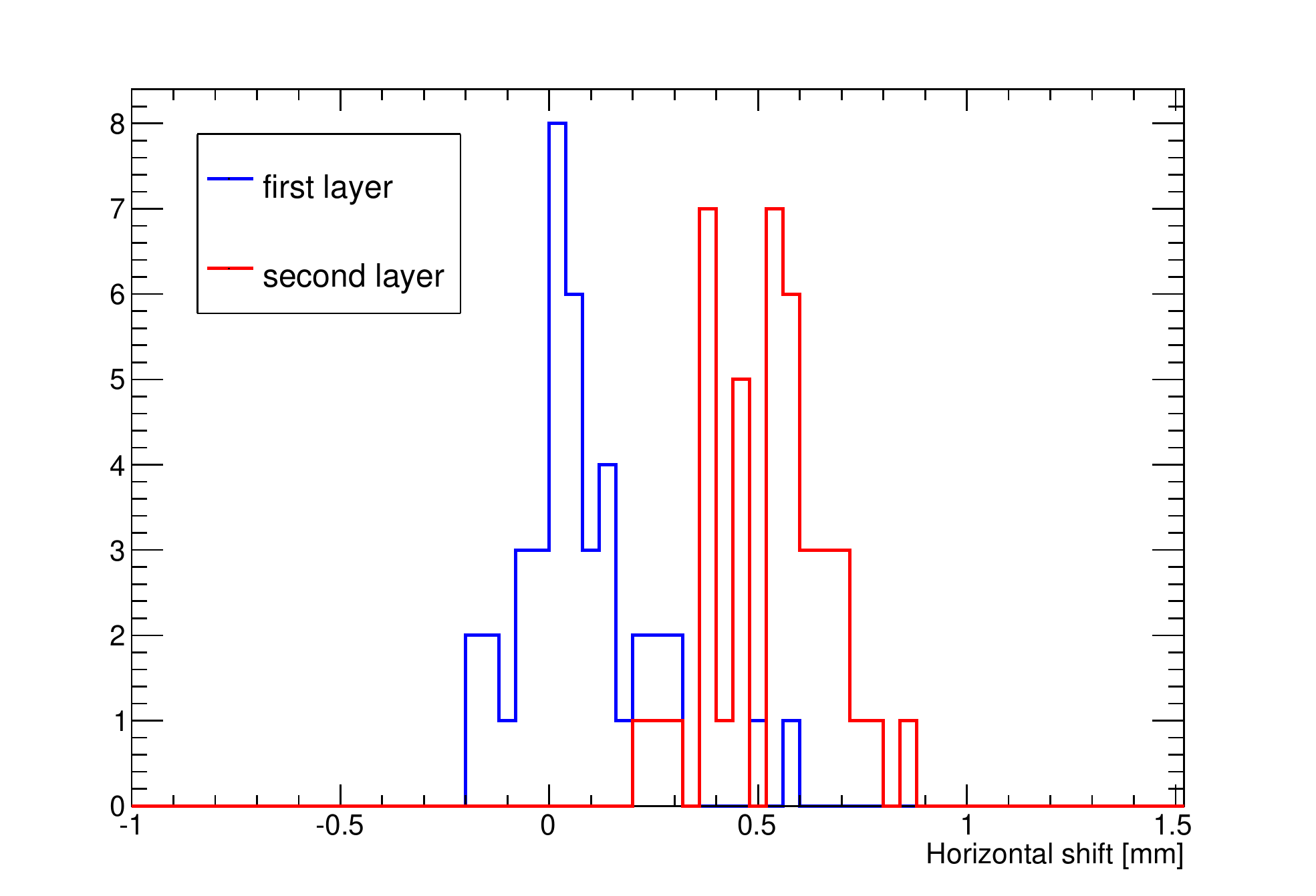}
\includegraphics[width=4.5cm]{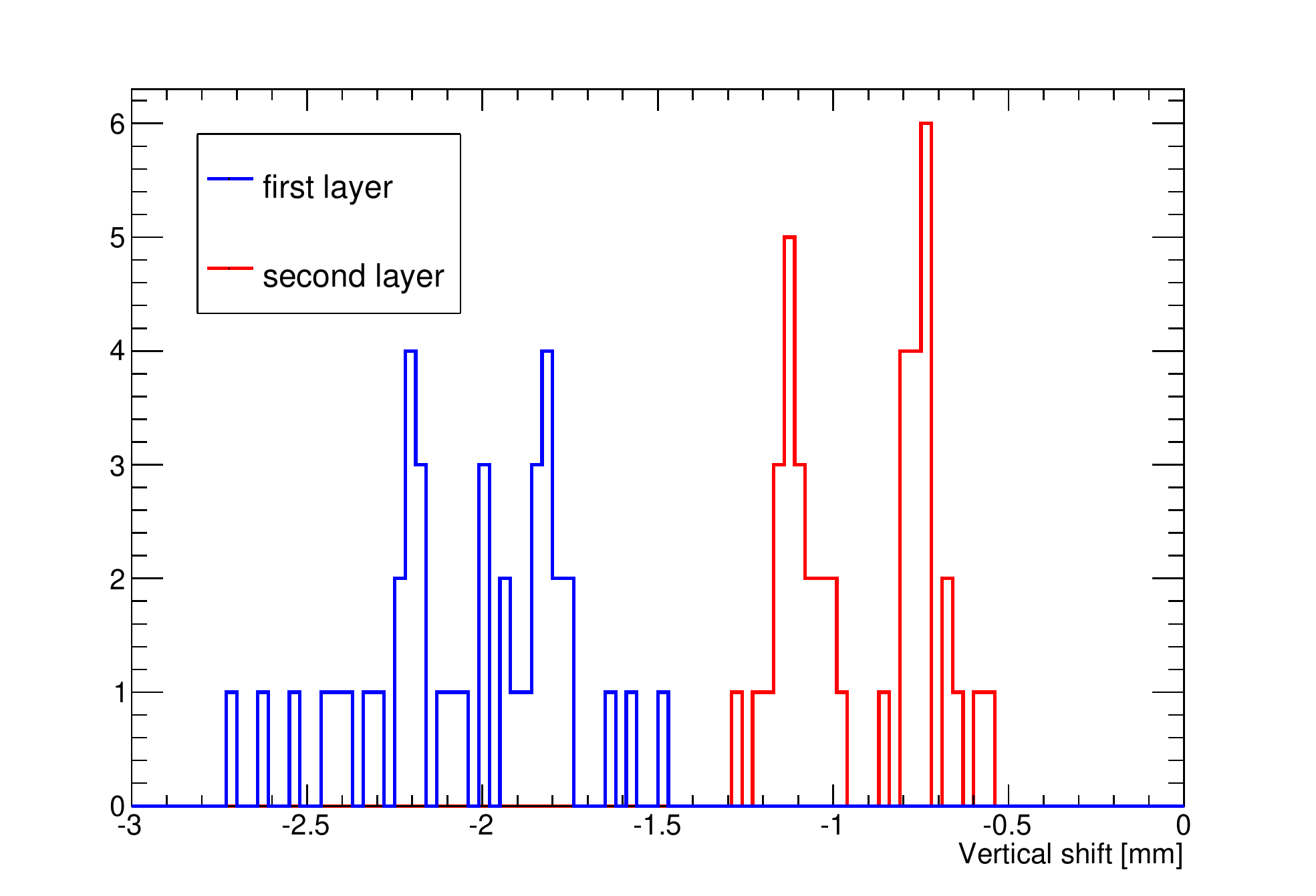}
\includegraphics[width=4.5cm]{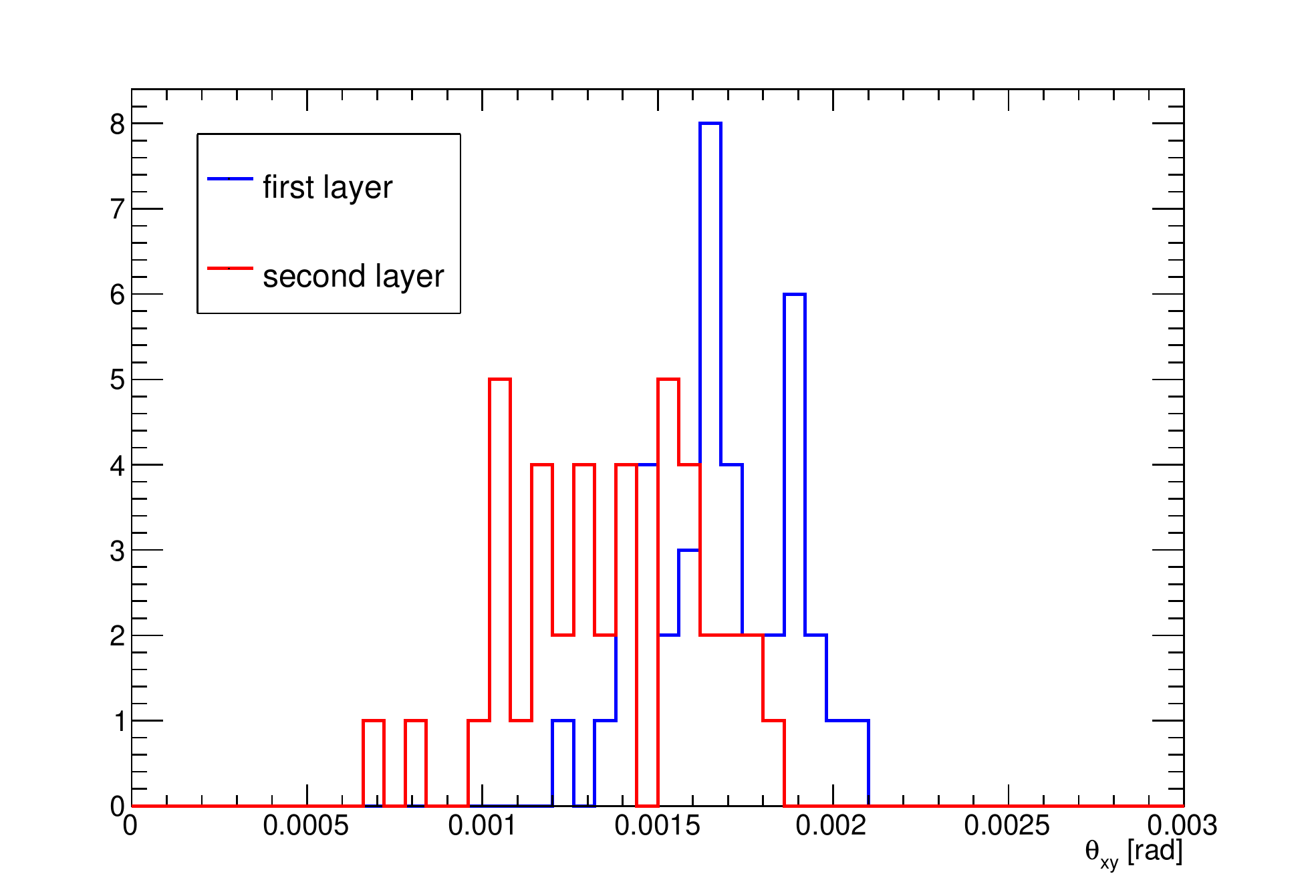}
\caption{Distributions of the alignment parameters, three columns corresponding to $H$ (left), $V$ (middle), and $\theta_{xy}$ (right). The blue and red histograms correspond to the first and second layer respectively.}
\label{fig6}
\end{figure}

The alignment parameters of the PSD will be used in the DAMPE's fight data reconstruction during its whole lifetime. Considering that plastic material changes its geometry heavily depending on temperature (Zhang et al. \cite{yp2017}, Li et al. \cite{Yao17}), we also studied the stability of the alignment parameters with time, since the Sun light angle changes seasonally causing the temperature variation. Technically, we divided one year's data into 4 groups, with a step of 3 months. MIP events in different time range can be used to get the alignment parameters, and the variations of these 4 groups are shown in Fig. \ref{fig5}. Almost all alignment parameters of PSD bars change only slightly except for the few bars located in the edge of PSD detector. This edge fluctuation is caused by lower statistics of corner segments, due to the lower geometrical acceptance of the BGO trigger for the border events. Fig. \ref{fig6} shows the alignment parameters after applying our method: the horizontal shift is relatively small, the two layers of PSD are shifted up, the rotation is counterclockwise, the vertical shift and rotation in xy plane are dominant. 

\begin{figure}
\includegraphics[width=4.5cm]{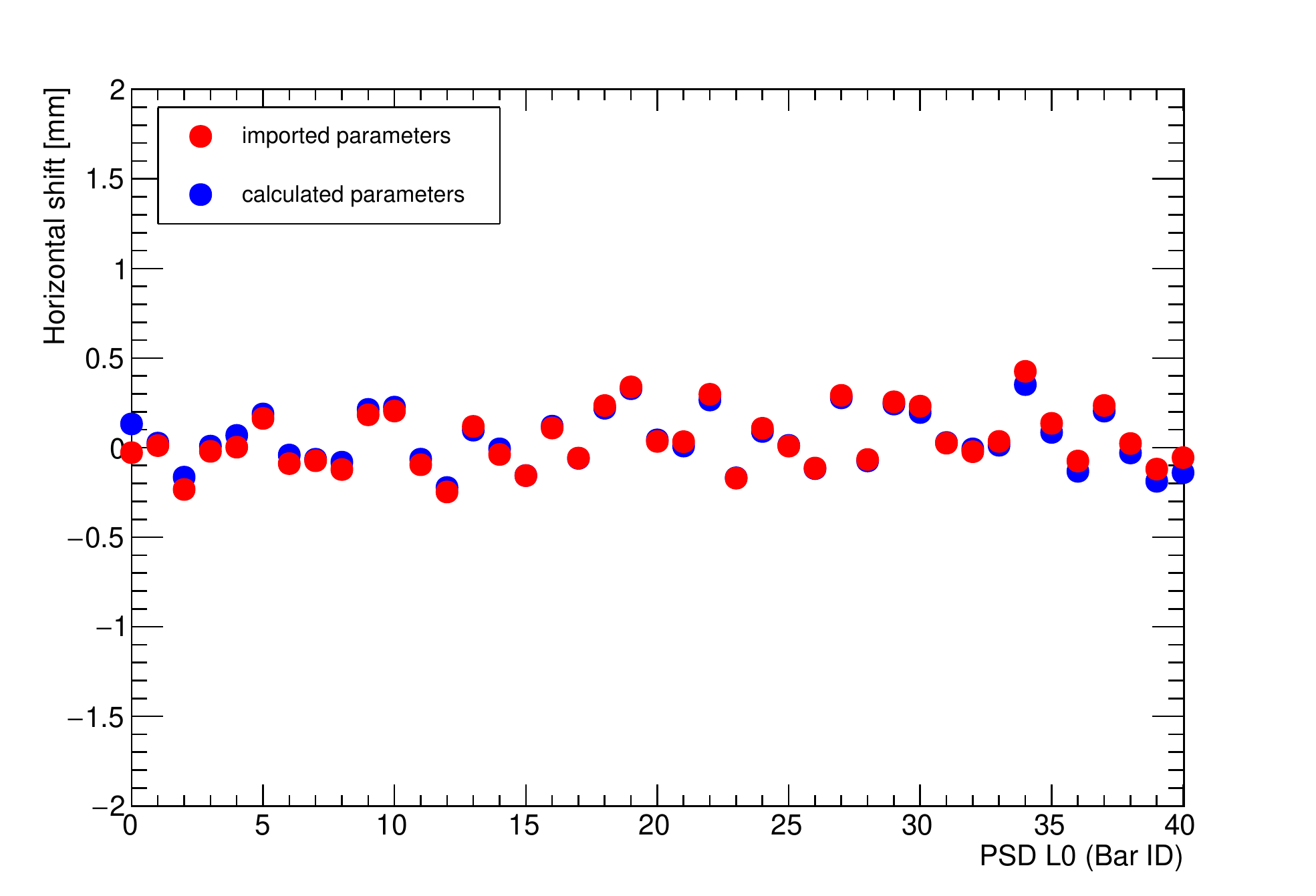}
\includegraphics[width=4.5cm]{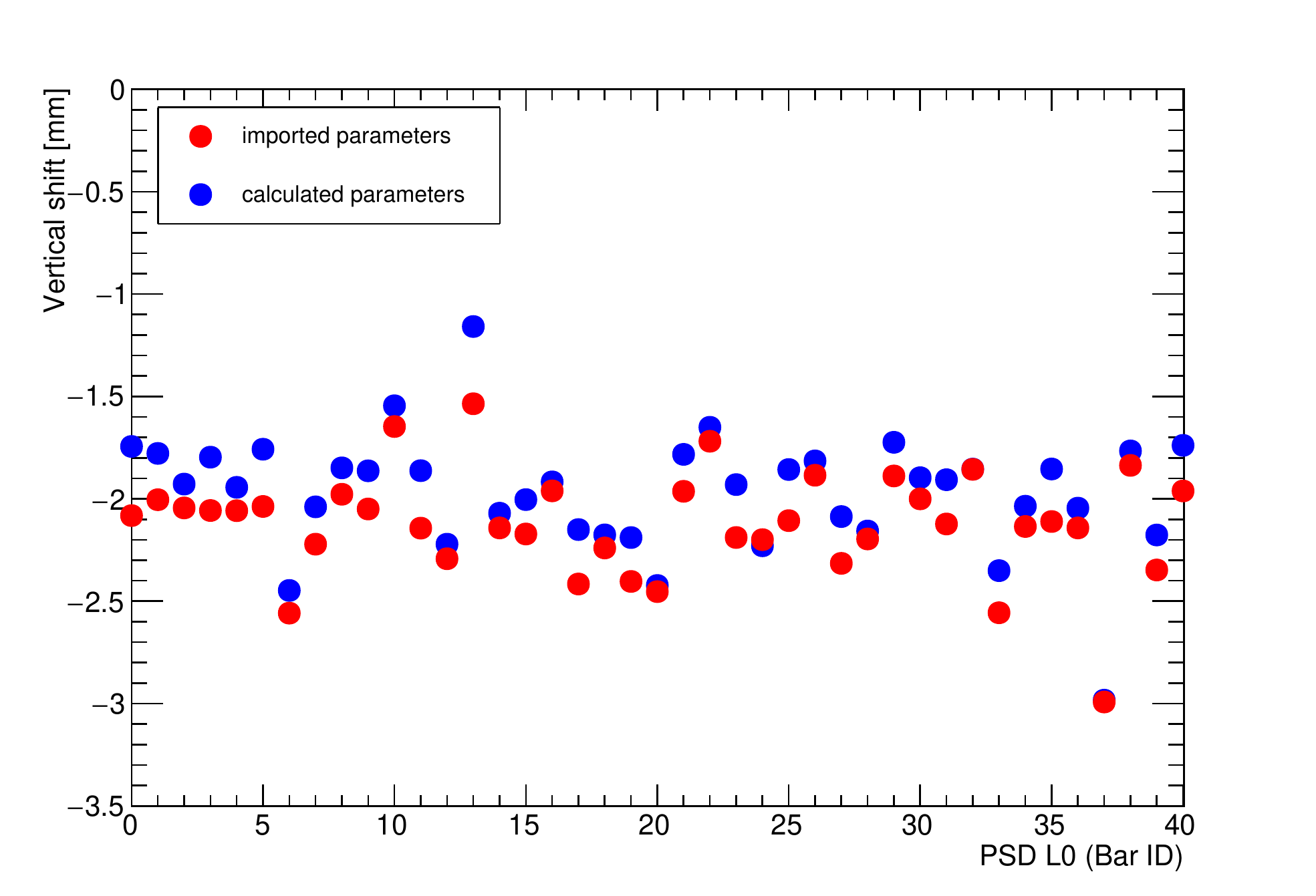}
\includegraphics[width=4.5cm]{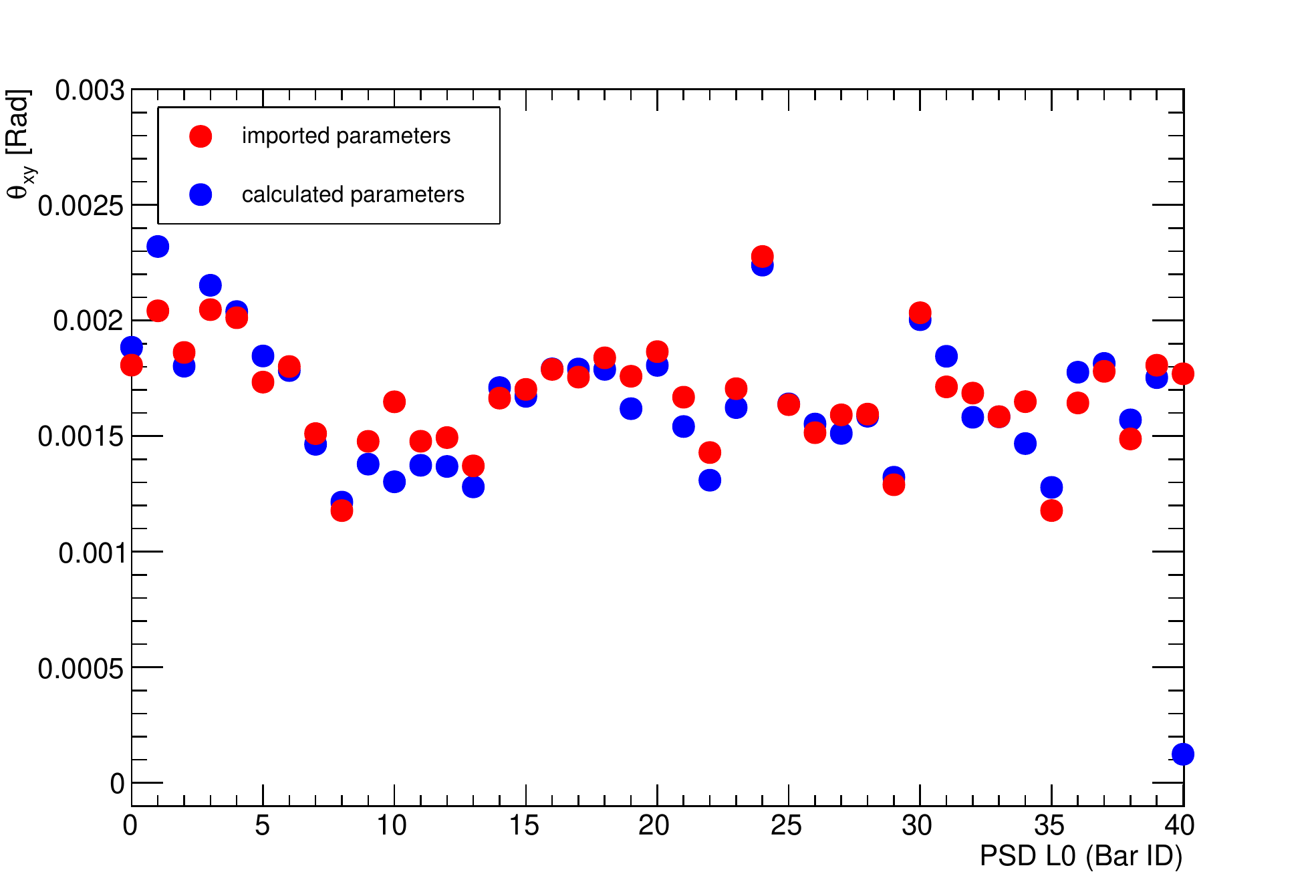}\\
\includegraphics[width=4.5cm]{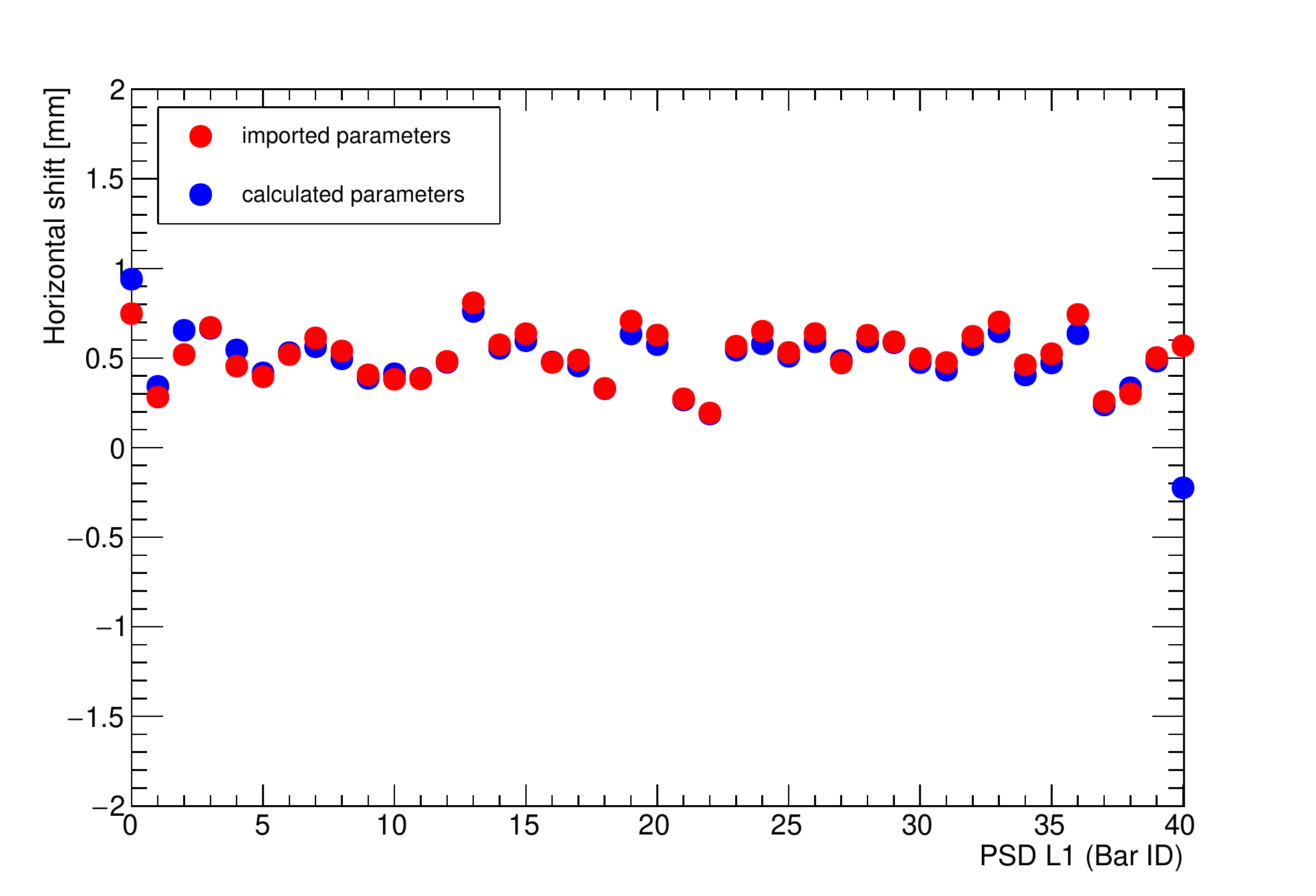}
\includegraphics[width=4.5cm]{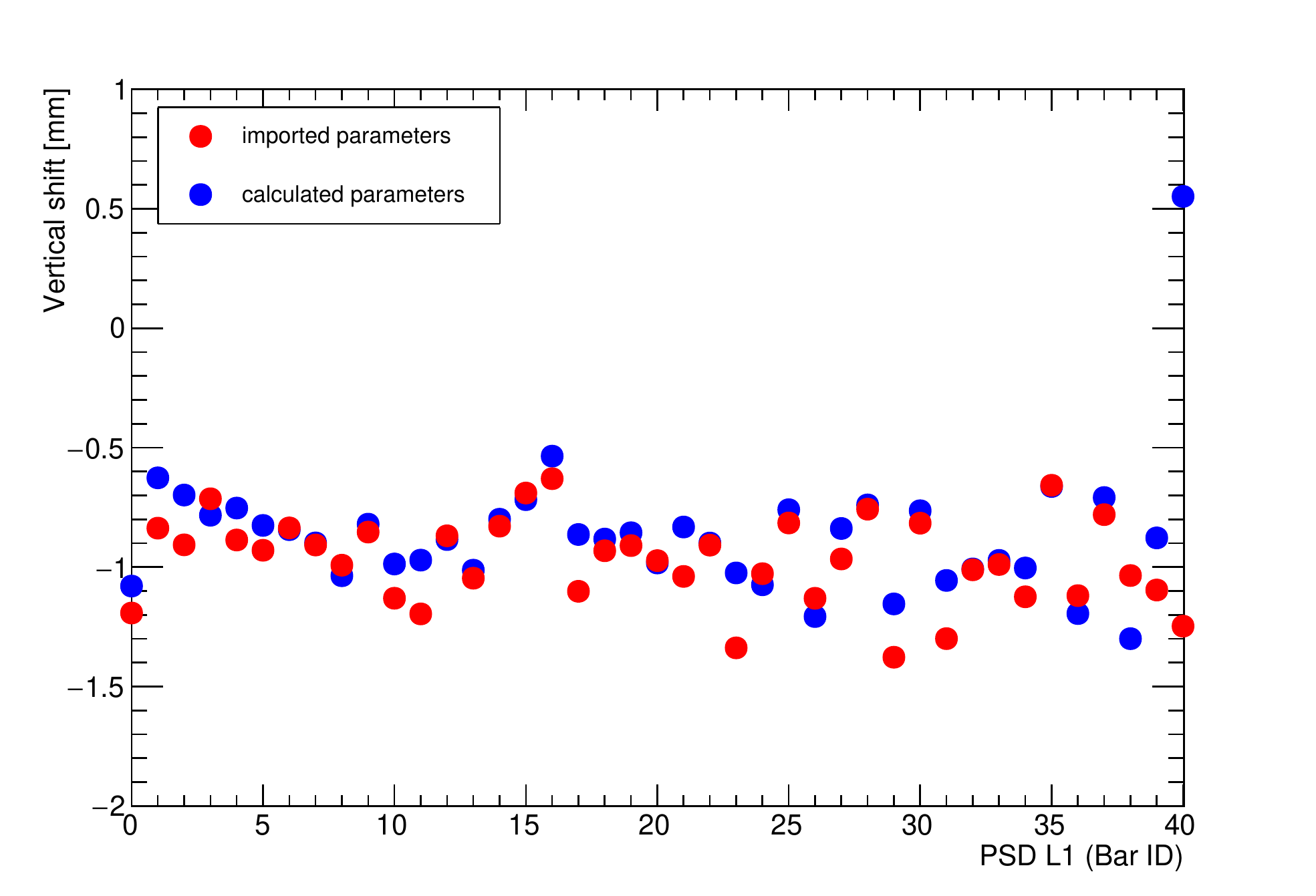}
\includegraphics[width=4.5cm]{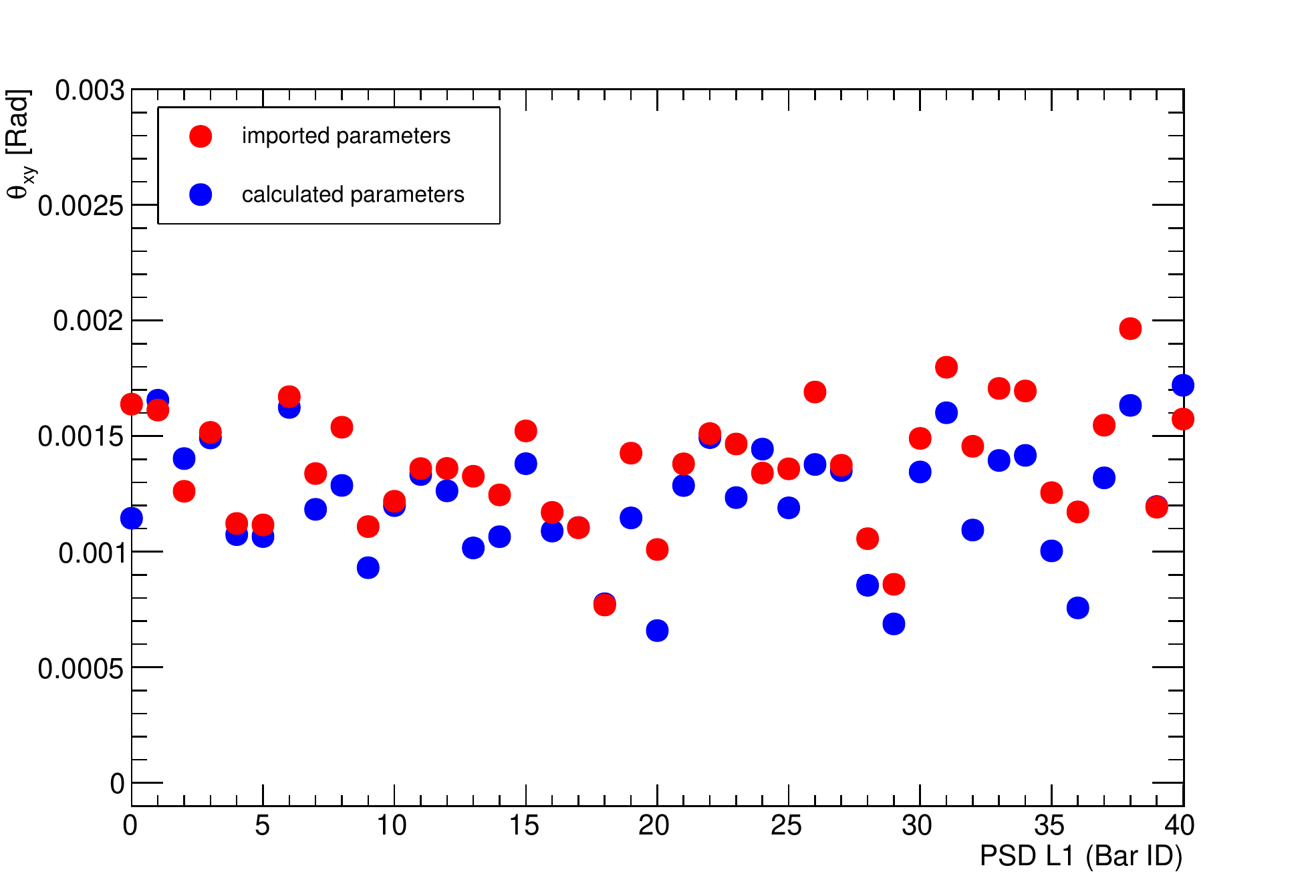}
\caption{Validation of the alignment with muon Monte-Carlo samples: red dots correspond to the misalignment parameters imported to the samples, while the blue dots are the alignment parameters calculated for these samples using our method. The first and second rows correspond to the first and second PSD layers respectively. The left, middle and right columns correspond to $H$, $V$ and $\theta_{xy}$ respectively.}
\label{fig7}
\end{figure}

To validate the alignment procedure, we manually import position shifts and rotations that come from the real geometry to the PSD geometry in Geant4 Monte Carlo simulation, then the same alignment method is applied to the misaligned Monte Carlo sample. In Fig. \ref{fig7} we show all the alignment variables $H$, $V$, and $\theta_{xy}$ extracted for each PSD bar. There is an overall good agreement between initial values and calculated ones.

\begin{figure}
\includegraphics[width=4.5cm]{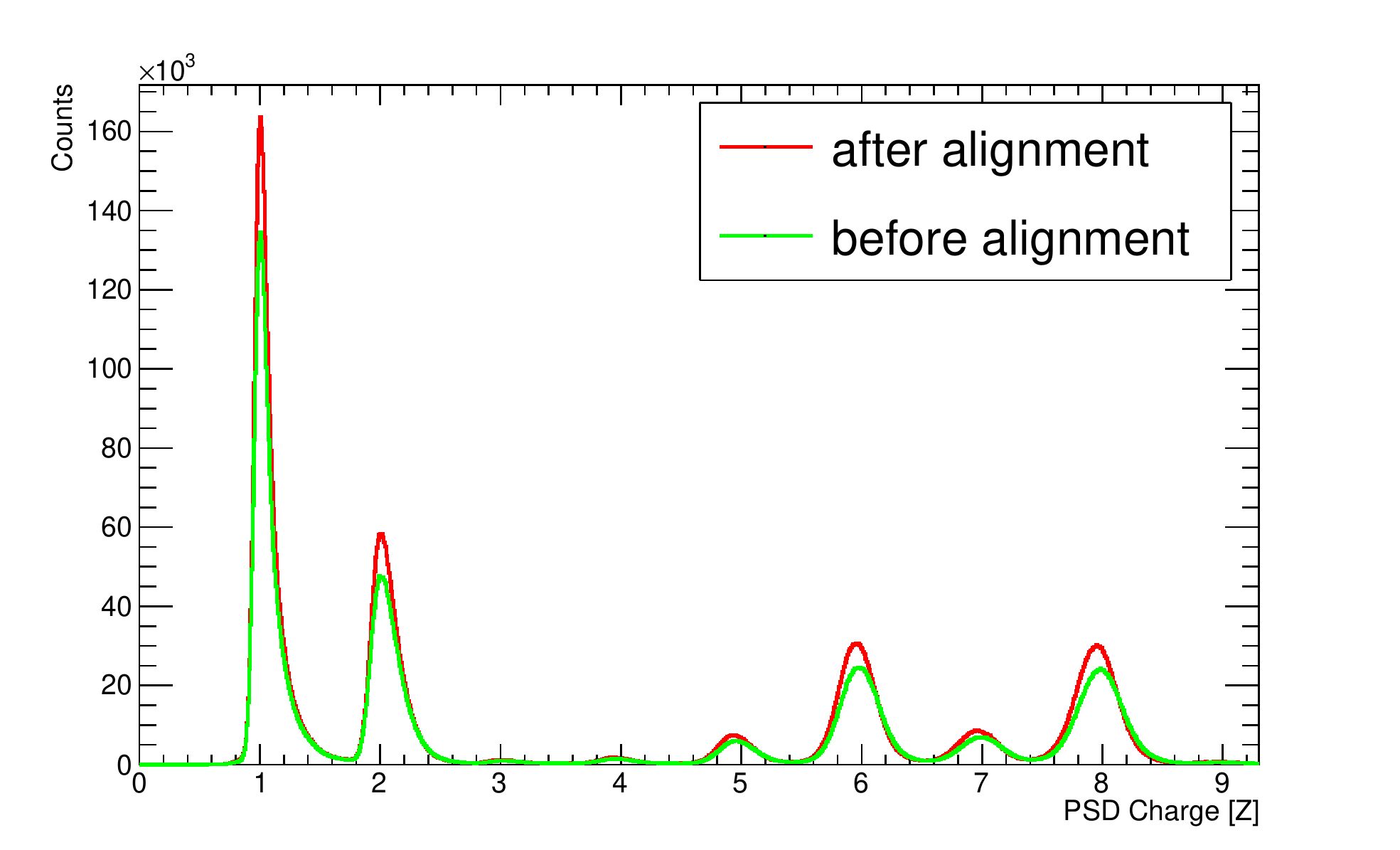}
\includegraphics[width=4.5cm]{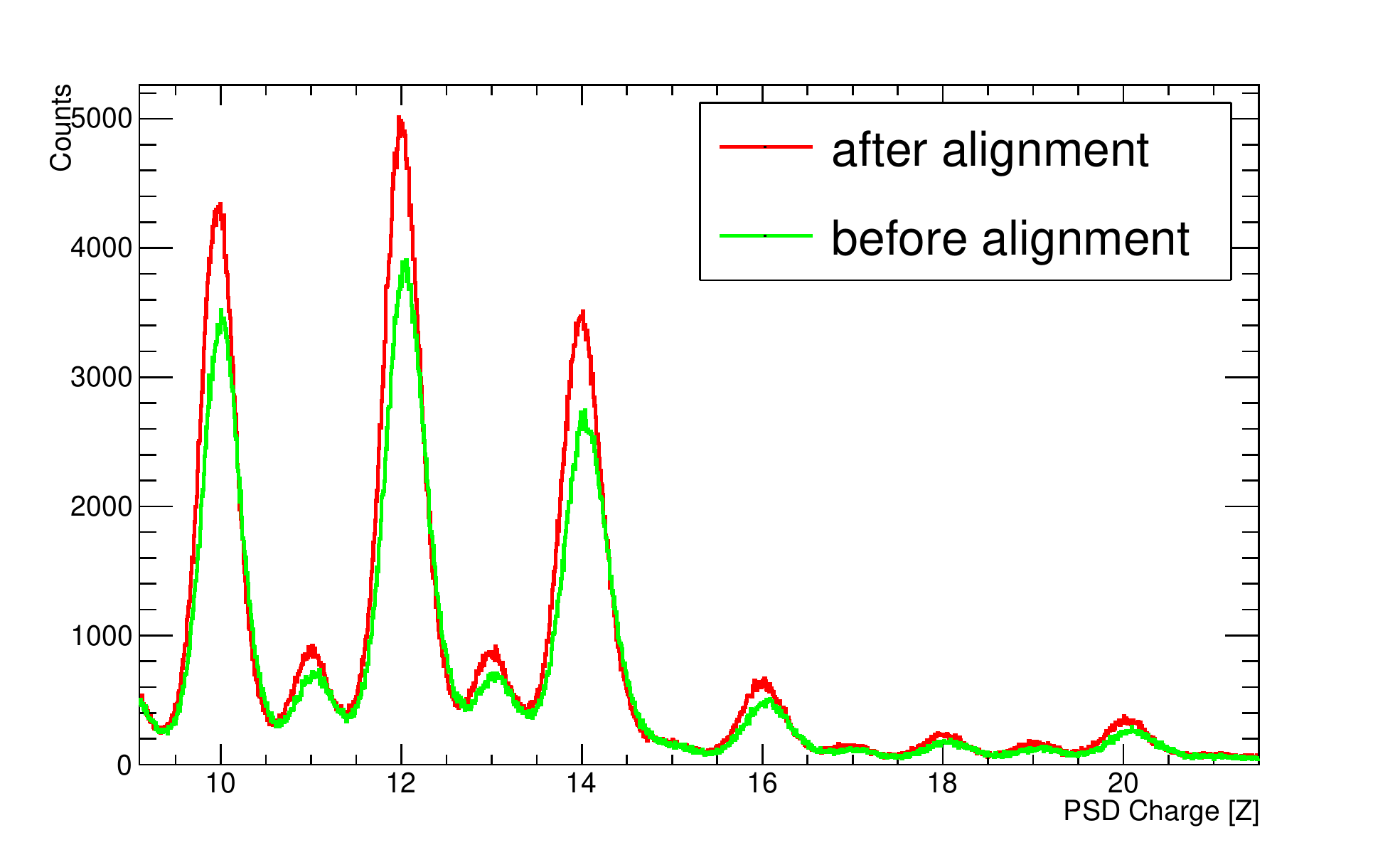}
\includegraphics[width=4.5cm]{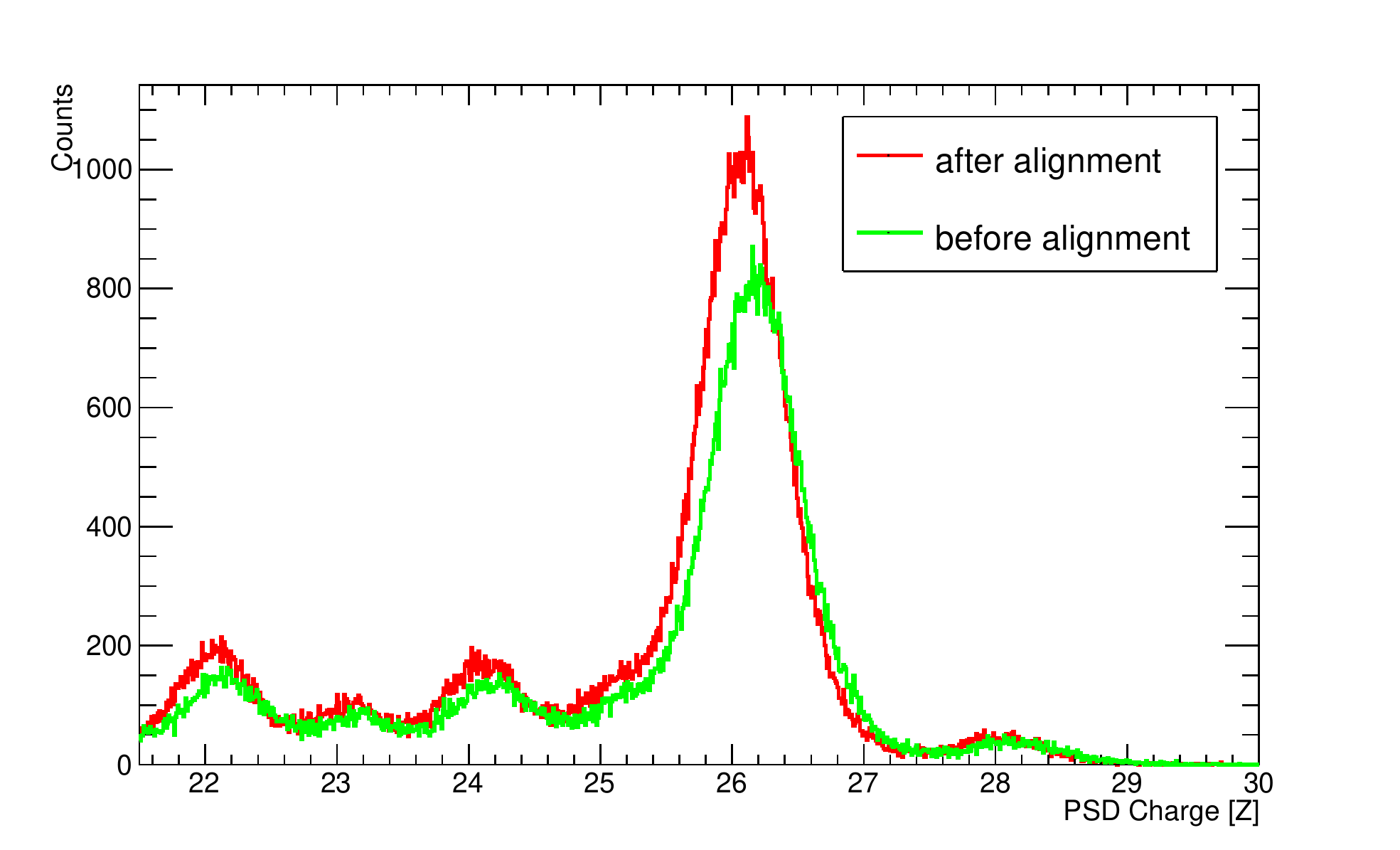}
\caption{The Comparison of PSD Charge spectrum: the red line is PSD charge with alignment correction, the green one shows the charge without alignment correction. One can see the significant improvement of charge resolution, which is particularly important for the physics analysis of cosmic rays}
\label{fig8}
\end{figure}

\begin{figure}
\includegraphics[width=4.5cm]{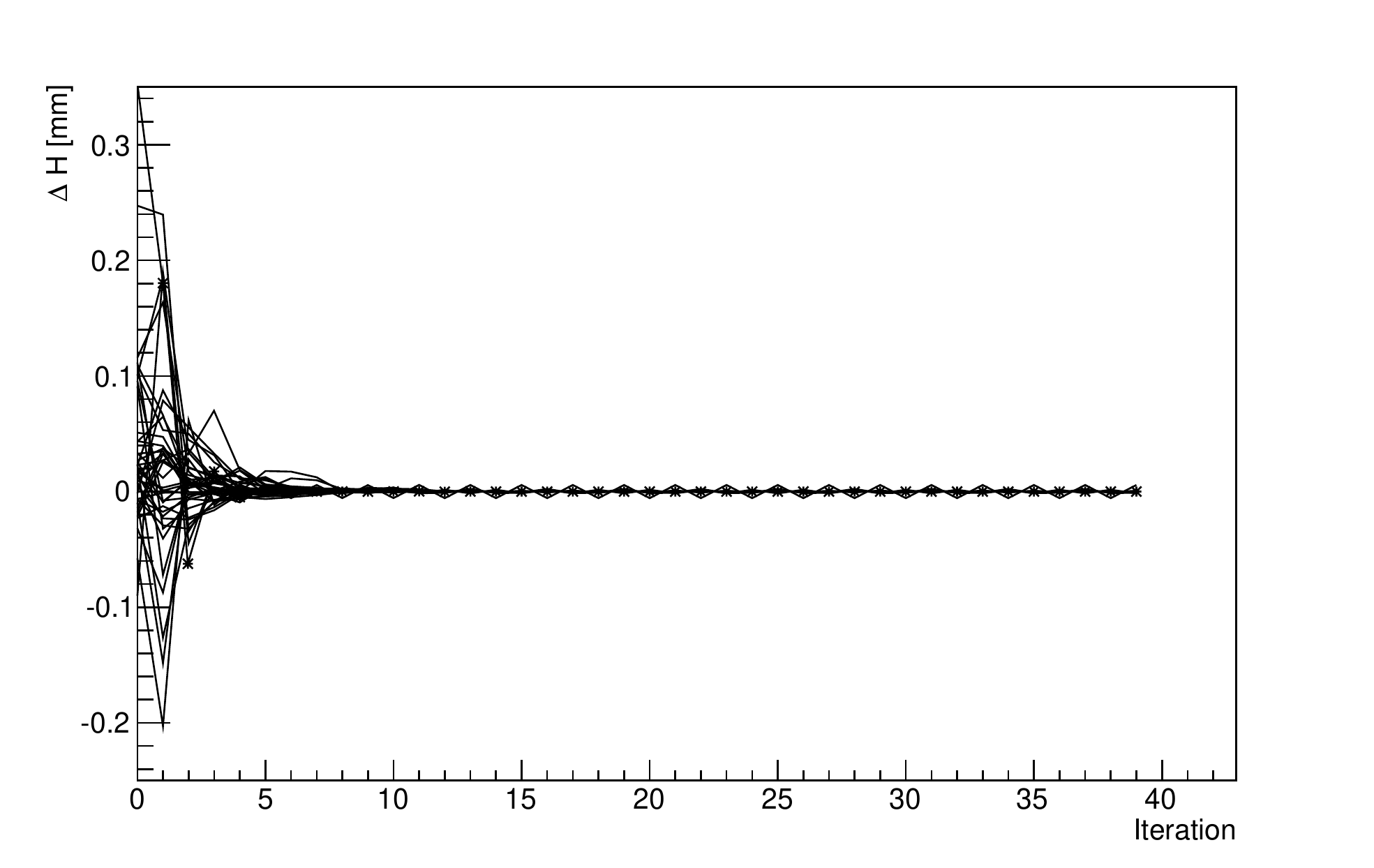}
\includegraphics[width=4.5cm]{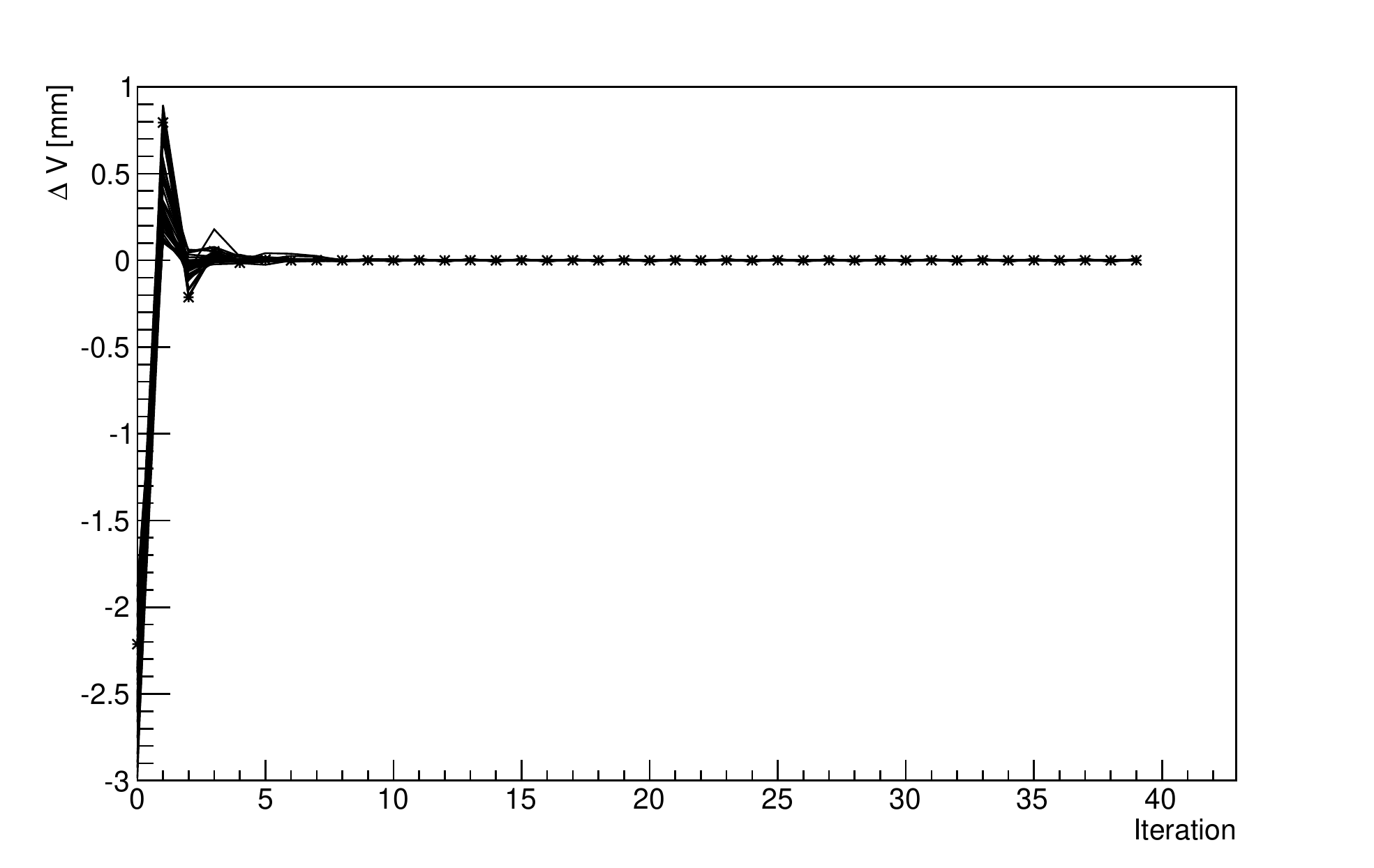}
\includegraphics[width=4.5cm]{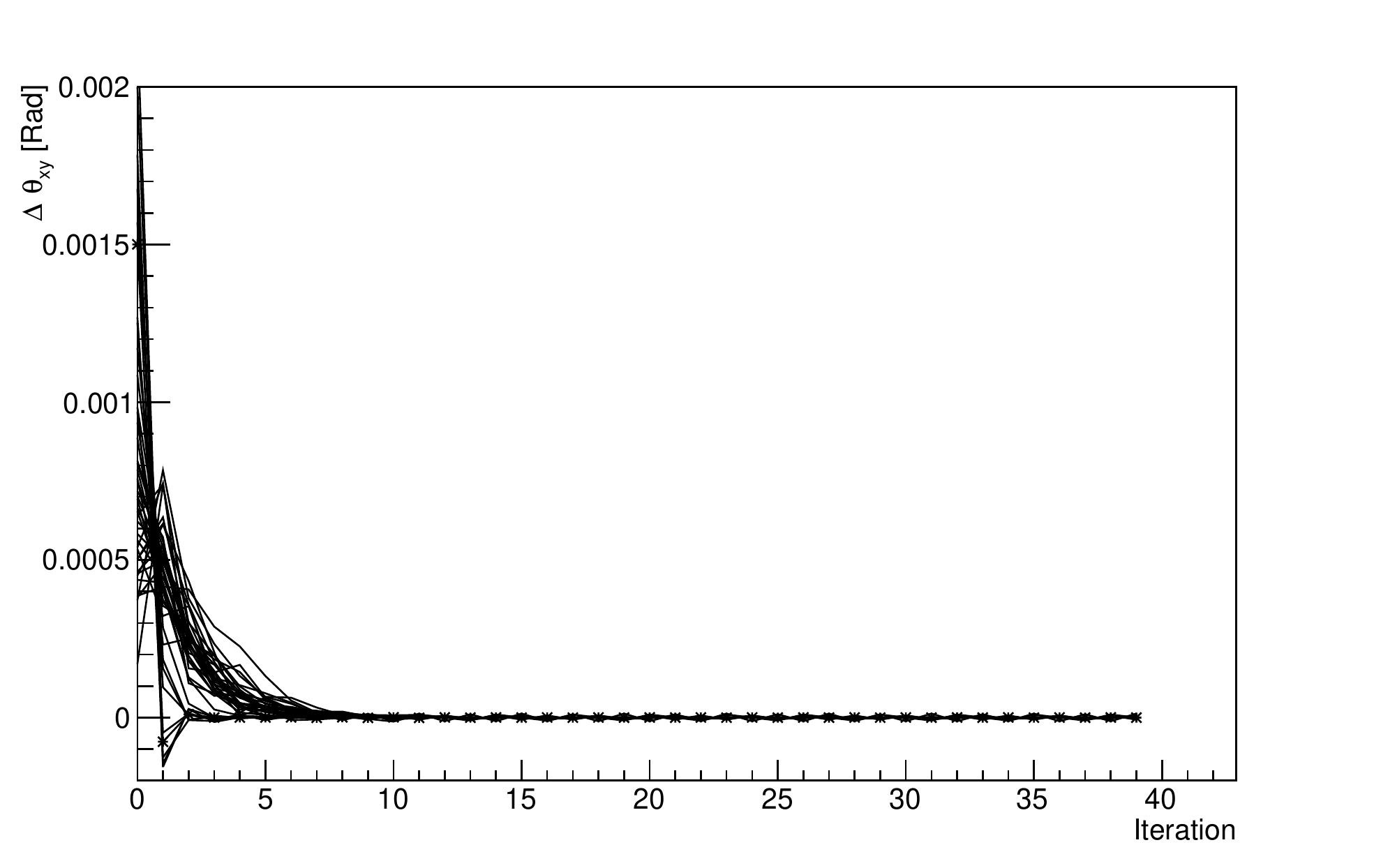}\\
\includegraphics[width=4.5cm]{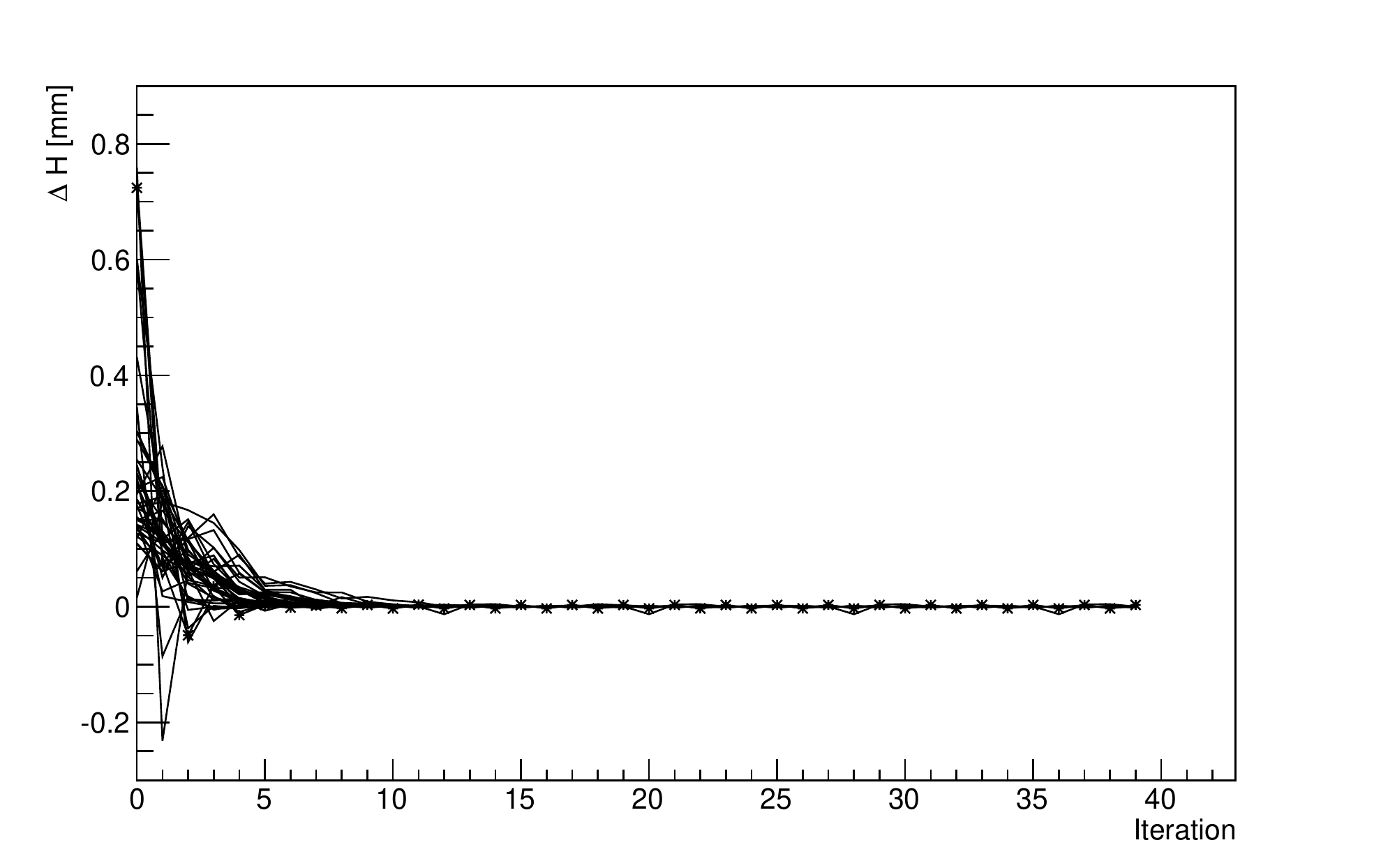}
\includegraphics[width=4.5cm]{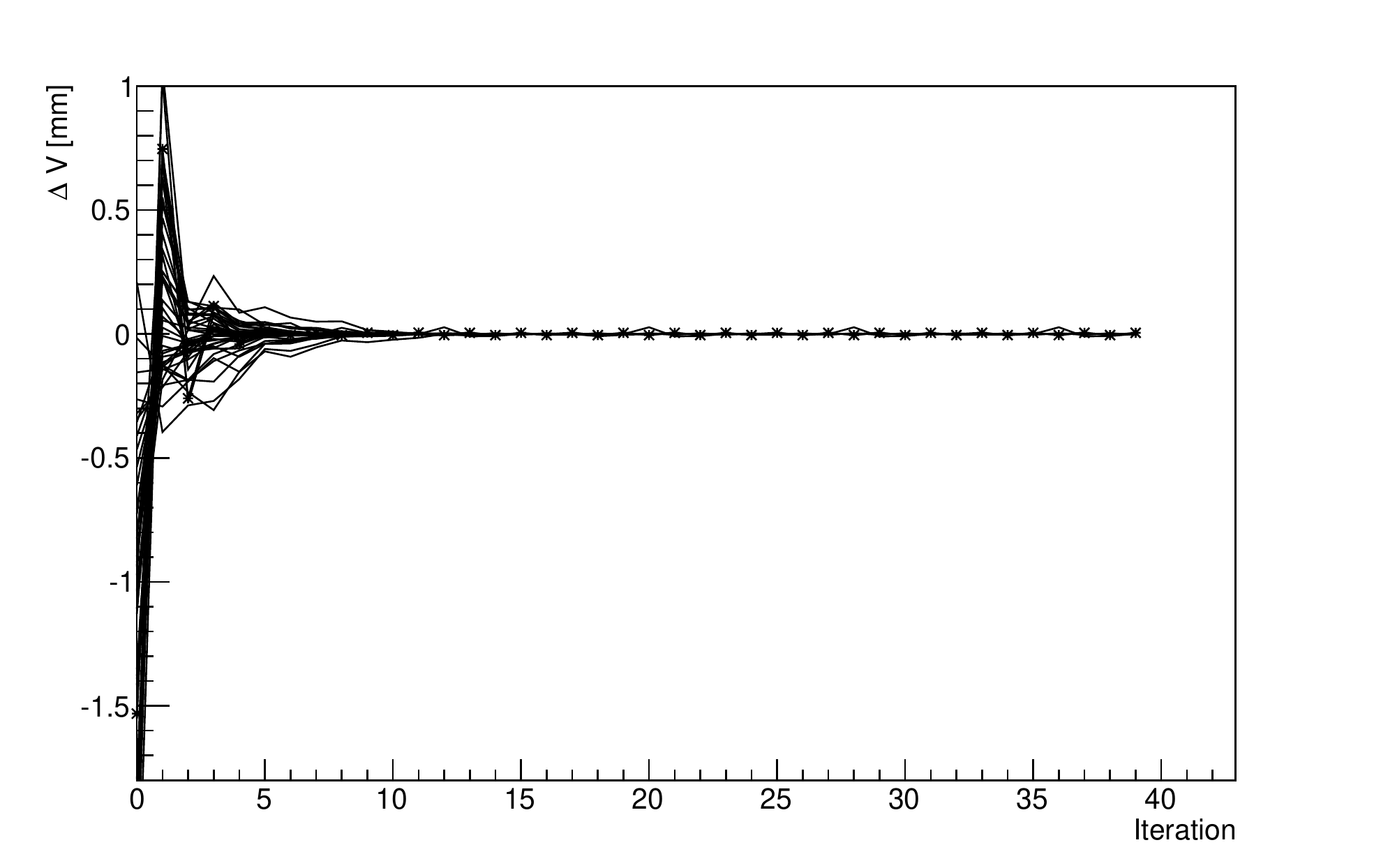}
\includegraphics[width=4.5cm]{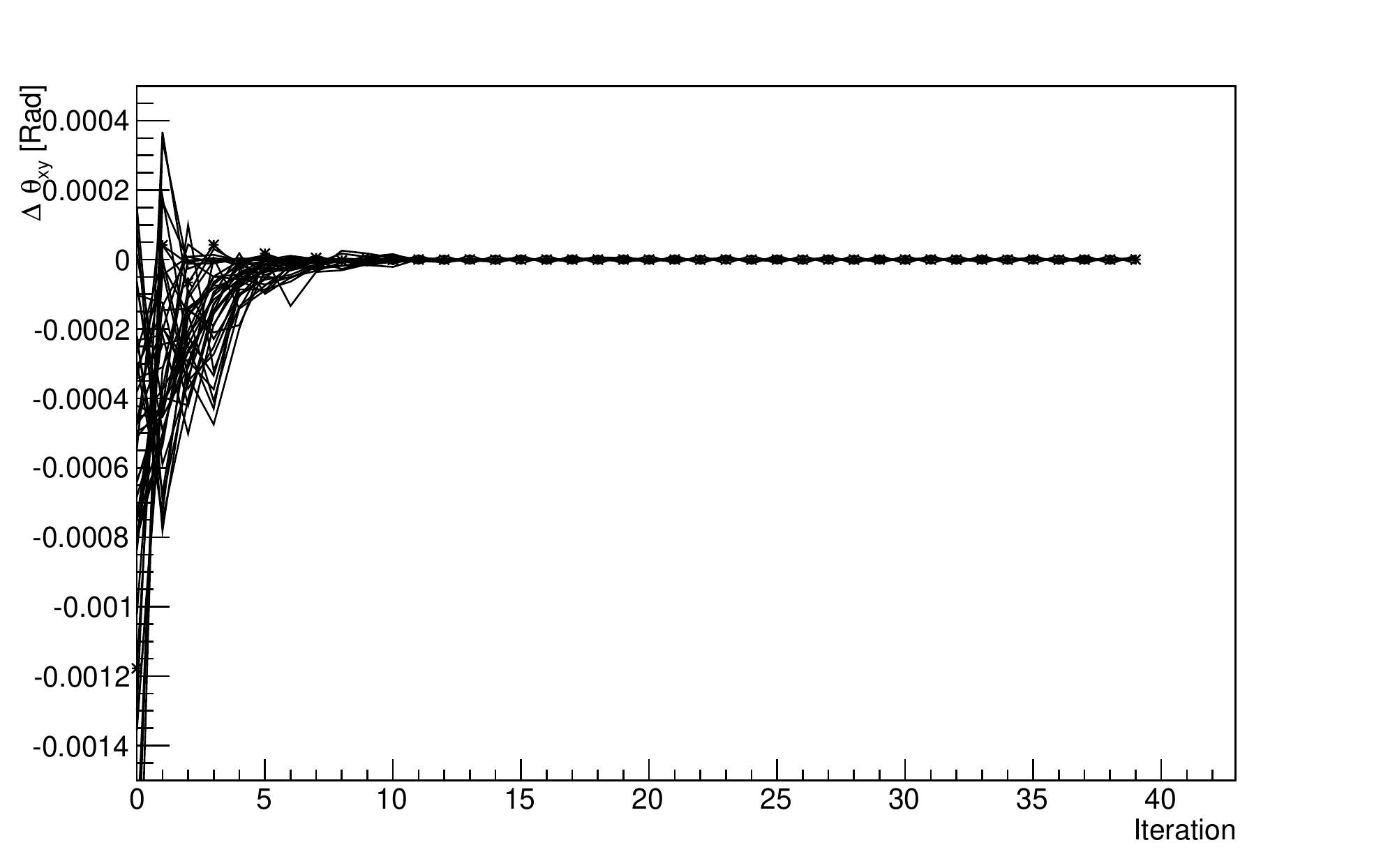}
\caption{Convergence of parameters variations with iteration. The first and second rows show convergence trend of the first PSD layer and the second PSD layer, respectively. Left, middle and right plots correspond to $H$, $V$, and $\theta_{xy}$ respectively}
\label{fig9}
\end{figure}

Based on DAMPE first-year data, we reconstruct the charge spectrum from H to Fe ,seen Fig. \ref{fig8} (Dong et al. \cite{dong18}), where the blue line is charge spectrum before the alignment, and the red one is after the alignment.  As seen from the figure, the charge resolution for all nuclei improves significantly especially for high abundance elements like H, He, C, N, O and Fe. Quantitatively, we summarize the charge resolution for several nuclei in Table. \ref{mbh}. After the alignment, charge resolution is improved by $4-8\%$.

\section{Result and possible application}
\label{sec4}

We presented the method to perform the DAMPE PSD detector alignment. Our main goal is to obtain the real PSD detector geometry information, which will be then applied to the DAMPE orbit data analysis. In particular, a precise PSD alignment is crucial for the measurement of cosmic ray nuclei flux. With the help of $\eta$ distribution of middle and corner events, shifts and rotations of PSD bars can be extracted natively and then integrated into the  designed PSD geometry. 
 After performing the alignment, we compare the $\eta$ distribution both for the middle and the corner events, before and after the alignment. Thanks to the alignment procedure, the apparent distortion of the $\eta$ distribution for the corner events is eliminated, proving a high efficiency of the alignment method. Due to the improved charge resolution after the PSD alignment, the presented result is expected to reduce significantly the systematic uncertainty of cosmic ray nuclei flux measurement.

Our alignment method is significantly different from the traditional detector alignment used in STK alignment (e.g. Tykhonov. et al. \cite{adr2018}). Traditionally, a precise track can be reconstructed and the residual between expected and calculated position can serve as a good quantity for performing the alignment. This traditional method can not be easily applied in case of PSD, since PSD bars have considerable size. On the other hand, the amplitude of MIP signal reflects the path length of a particle track and can be precisely measured. Therefore, we show that the signal amplitude is a quantity which can be successfully used for performing the alignment. 
 Certainly, our alignment method requires a precise track first. Finally, we believe that our methodology can be successfully applied for the alignment of other large-scale detector units.

\begin{acknowledgements}
We thank A. Tykhonov for helpful feedback on previous drafts. This work was funded by the National Key Program for Research and Development (No. 2016YFA0400200), the Strategic Priority Research Program of the Chinese Academy of Sciences (No. XDB23040000) and the National Natural Science Foundation of China (NSFC) (Grant Nos. 11773086, U1738205, U1738127, 11673021, 11673047, 11673075, 11643011, 11773085, U1738207, U1738138, U1631111, U1738129 and 11703062).
\end{acknowledgements}


\begin{table}
\bc
\begin{minipage}[]{100mm}
\caption[]{Improvement of charge resolution after applying the PSD alignment correction. The charge resolution corresponds to either a width of Landau fit (Z=1,2) or $\sigma$ of Gaussian fit (Z$>$2).\label{mbh}}\end{minipage}
\setlength{\tabcolsep}{2.5pt}
\small
 \begin{tabular}{ccccccccccccc}
  \hline\noalign{\smallskip}
H & He & Li & Be & B & C & O & Ne & Mg & Si  \\
 \hline\noalign{\smallskip}
0.037 & 0.056 & 0.126 & 0.124 & 0.138 & 0.156 & 0.202 & 0.239 & 0.254 & 0.286 \\
 \hline\noalign{\smallskip}
0.035  & 0.051 & 0.119 & 0.119 & 0.131 & 0.149 & 0.193 & 0.229 & 0.240 & 0.274 \\
 \hline\noalign{\smallskip}	
5.4\% & 8.9\% & 5.5\% & 4.0\% & 5.1\% & 4.5\% & 4.5\% & 4.2\% & 5.5\% & 4.2\% \\
  \noalign{\smallskip}\hline
\end{tabular}

\ec

\end{table}

\label{lastpage}

\end{document}